\documentclass[preprint]{elsarticle}

\usepackage{color} 
\usepackage{xcolor}

\usepackage{geometry} 
\usepackage{amsthm}
\usepackage{amsmath}
\usepackage{amssymb}
\usepackage{dsfont}
\usepackage{amsbsy}

\usepackage[resetlabels]{multibib}
\newcites{pri}{References}
\newcites{sec}{References}

\usepackage{algorithm}
\usepackage{algorithmic}
\usepackage[algo2e]{algorithm2e}

\usepackage{pdfpages} 

\newcommand{\review}[1]{\textcolor{black}{#1}} 

\newtheorem{mydef}{Definition}

\newtheorem{property}{Property}

\newcommand{\E}{\mathbb{E}}
\newcommand{\Var}{\mathbb{V}\mathrm{ar}}

\usepackage{lineno,hyperref}

\journal{Journal of \LaTeX\ Templates}

\date{December 4, 2020}
\begin{document}

\begin{frontmatter}

\title{Functional principal component analysis for global sensitivity analysis of model with spatial output.}

\author[1]{T.V.E. Perrin}
\address[1]{Mines Saint-\'{E}tienne, Univ. Clermont Auvergne, CNRS, UMR 6158 LIMOS,\\ Institut Henri Fayol, F-42023, Saint-\'{E}tienne, France}

\author[2]{O. Roustant}
\address[2]{Institut de Math\'ematiques de Toulouse, Universit\'e de Toulouse, INSA,\\ 135, Avenue de Rangueil, 31077 Toulouse Cedex 4, France}

\author[3]{J. Rohmer}
\address[3]{BRGM, 3 av. Claude Guillemin, BP 36009, 45060 Orléans Cedex 2, France}

\author[4]{O. Alata}
\address[4]{Lab. Hubert Curien, UMR CNRS 5516, UJM-Saint-Étienne, IOGS, Univ. de Lyon, 42023, Saint-Étienne, France}

\author[5]{J.P. Naulin}
\address[5]{CCR, 157 Boulevard Haussmann, 75008 Paris, France}

\author[3]{D. Idier}
\author[3]{R. Pedreros}

\author[5]{D. Moncoulon}
\author[5]{P. Tinard}

%
%

\begin{abstract}
Motivated by risk assessment of coastal flooding, we consider time-consuming simulators with a spatial output.
The aim is to perform sensitivity analysis (SA), quantifying the influence of input parameters on the output.
There are three main issues. First, due to computational time, standard SA techniques cannot be directly applied on the simulator.
Second, the output is infinite dimensional, or at least high dimensional if the output is discretized. Third, the spatial output is non-stationary and exhibits strong local variations.\\
We show that all these issues can be addressed all together by using functional PCA (FPCA). 
We first specify a functional basis, such as wavelets or B-splines, designed to handle local variations. 
Secondly, \review{we select the most influential basis terms,
either with an energy criterion after basis orthonormalization, or directly on the original basis with a penalized regression approach}.
Then FPCA \review{further} reduces dimension by doing PCA on the \review{most influential} basis coefficients, with an ad-hoc metric. 
Finally, fast-to-evaluate metamodels are built on the few selected principal components. They provide a proxy on which SA can be done.
As a by-product, we obtain analytical formulas for variance-based sensitivity indices, generalizing known formula assuming orthonormality of basis functions.
\end{abstract}


\end{frontmatter}

%
%
\setcounter{section}{0} 
\section{Introduction}
\label{sec:introduction}
Coastal flooding may lead to major natural disasters in coastal regions \cite{chaumillon2017storm}, as exemplified by several recent events like cyclone Irma  in 2017 or Hurricane Sandy in 2012. In France, the last major event is Xynthia storm that induced 53 deaths, 79 injured people and 2.5 billions euros of damages, whose 700 million euros for coastal flooding (see e.g., \cite{naulin2016estimation},\cite{Xynthia2011}).
The technical pillar of any flooding risk assessment is the capability for accurate and robust predictions of the inland consequences (i.e. water levels at the coast, flood spatial extent, etc.) given any offshore meteo-oceanic conditions (like surge peak, tide peak, storm duration, wave characteristics, etc.). This can be done using high-resolution hydrodynamic numerical models (i.e. simulators). In the current study, we consider the \review{spatial distribution of the maximum value of the water depth (calculated over the time duration of a given storm event) as a typical indicator of flooding: this is the output of the considered numerical simulator. The inputs are offshore meteo-oceanic conditions, and correspond in our case to the main characteristics describing the time evolution of the surge and of the tide, i.e. the peak magnitude, the difference between both signals, the duration of the surge, etc: these} are associated to high degree of uncertainty. Therefore, we aim at evaluating the influence of these uncertainties on the \review{spatialized maximum water depth}.\\


To do so, \review{we are interested in non-intrusive methods: the simulator is considered as a black box which can only be evaluated. In this context, several} sensitivity analysis techniques have been proposed 
(see, e.g., \cite{iooss2015review, saltelli2008global}). 
However, there are two main issues. First, Monte Carlo methods commonly used to estimate sensitivity indices of each input parameter, require a large number of simulator runs (more than $10^4$). Hence, they are hardly applicable directly on the simulator, which typically presents large computation time cost for a single run (of several minutes, even hours).
Second, the output is functional: the maximum water depth is a function of the location. In practice, the locations are discretized, and the output is represented by a high dimensional vector of length equal to the number of pixels. Depending on the processes involved in the flooding (overflow, wave-induced overtopping, coastal defences' breaching, see an exhaustive overview by \cite{chaumillon2017storm}), the required level of discretization can be very fine (down to a few meters). This might add difficulty for sensitivity analysis by imposing to manipulate vector of high dimension (typically above 10,000 : see section \ref{sec:StudyCase}).\\

In this context, a standard methodology fixes these two problems in the following way
(\review{see e.g. \cite{chen2011efficient, marrel2010global, jia2013, marrel2015,li2020, Ma2019}}).
First, the output dimension is reduced, most often by principal component analysis (PCA)
or by using functional basis decomposition (using e.g. on Fourier basis or wavelets).
This provides a lower dimensional output vector, formed by the largest components (PCA components or basis coordinates).
Second, a fast-to-evaluate proxy, called metamodel \review{or surrogate model}, is built on that vector. 
This is usually done by considering independently each coordinate as a scalar output \review{as in \cite{chen2011efficient}}.
Among all metamodels (e.g. linear regression, neural networks, etc.),  we select the Gaussian process (GP) regression model \cite{williams2006gaussian}, because they provide both an interpolation of the data and an uncertainty at unknown area; moreover,
the method is parameterized by a covariance function (or kernel), 
which makes it flexible, and allows to exploit expert knowledge.\\


\review{Unfortunately,} PCA treats each coordinate independently and misses the spatial dependence.
Furthermore, the output of the simulator exhibits strong local variations \review{(this is illustrated by Figure \ref{fig:CoastalModelOutput}), and further discussed in our application case in Sect. \ref{sec:TestCaseFlooding}}, 
meaning that the water \review{depth} is not a smooth function of the location.
As a result, even with suitable functional bases such as wavelets, 
a large number of coefficients, typically several hundreds, 
must be kept to get an accurate approximation. \review{This problem has clearly been highlighted in previous studies, e.g. \cite{marrel2010global, marrel2015}}. This weakens the benefits of reducing dimension.\\

To tackle this \review{issue}, we propose to use functional PCA (FPCA), a common technique in functional data analysis \cite{ramsay2006functional}. This is equivalent to performing PCA on the coefficients of a functional basis decomposition, with the metric given by the Gram matrix of basis functions. It can be used for popular bases, including Fourier, wavelets, and B-splines. Notice that for non-orthonormal bases such as B-splines, the PCA step uses a different metric than the usual PCA. 
\review{In addition, we add a preliminary selection step, by choosing the basis terms which are most influential, based on the energy decomposition. With these two ideas, 
the method is applicable for large dimensional vectors.
For instance, we can deal with maps made of several tens of thousands of pixels.
Furthermore, we cumulate the advantages of PCA and basis decomposition by accounting for spatial dependence of the output, 
which is ignored by standard PCA}, 
since functions are decomposed in a suitable functional space. 
Besides, by doing PCA in a second time, dimension reduction is ensured, 
even when a large number of basis coefficients must be kept: 
the final number of principal components is small. 
Finally, as remarked when doing PCA, building a metamodel independently for each coordinate has some sense, since the principal components are uncorrelated (though not necessarily independent).\\

The use of FPCA for sensitivity analysis has been proposed for instance by \cite{lamboni2011multivariate}.
There, orthonormal basis functions are obtained as eigenfunctions of a Hilbert-Schmidt operator associated to a covariance kernel, by Karhunen-Lo\`eve decomposition. 
In our approach, we define the functional basis first. In theory, the two approaches are equivalent, since a covariance kernel can be built from a predefined basis corresponding to its Karhunen-Lo\`eve decomposition. However, in practice, here, there is a clear advantage in defining the functional basis first, which is to deal with non stationarity without expert knowledge. Indeed, contrarily to usual kernels in RKHS which are guided by global regularity assumptions, several functional basis such as wavelet basis have been designed to fit functions with strong local variations.\\

As a second contribution, we give a closed-form expression for Sobol indices by using FPCA metamodels. As explained in the previous paragraph, the formula can be made equivalent to the expression found in \cite{lamboni2011multivariate} (these indices are named ``generalized sensitivity indices'') in the case of orthonormal basis functions, 
when using a kernel constructed from the basis functions. The formula that we obtain is also valid for non-orthonormal popular basis functions such as B-splines.\\


The paper is organized as follows. 
Section \ref{sec::background} introduces technical backgrounds on the different methods used in this study: 
functional PCA, global sensitivity analysis and GP regression models.
Section \ref{sec:metamodelsFPCA} presents our contribution for metamodelling with spatial output.
Section \ref{sec:GSIprop} extends the generalized sensitivity indices for non-orthonormal basis. The proposed procedure is applied on two case studies. The first one is an analytical case used to describe and illustrate the different steps of the proposed procedure (Section \ref{sec:TestCase}). The second one corresponds to a real case of coastal flooding on the french Atlantic coast (Section \ref{sec:StudyCase}).
Finally, Section \ref{sec:conclusion} discusses the main results and identifies potential lines for future works.

%
%
\section{Background}
\label{sec::background}

\subsection{Functional principal component analysis}
\label{sec::background::subsec::FPCA}
FPCA is widely used in Functional Data Analysis (FDA) to find the dominant
modes of variation in a set of functions, here 2-dimensional maps. 
These modes correspond to functions of a lower finite dimensional basis, where data can be represented. \review{They correspond to the eigenfunctions basis, also called principal components.}

\review{The theory of FPCA relies on the Karhunen-Lo\`eve (KL) decomposition of random fields.
In this framework, functions are viewed as realizations of a centered Gaussian random field $Y$ with covariance kernel $k$.
Under specific assumptions, $Y$ admits a KL decomposition of the form
$$ Y(x) = \sum_{n \in \mathbb{N}} \sqrt{\lambda_n} \epsilon_n \phi_n(x)$$
where the $\epsilon_n$'s are i.i.d. standard Normal random variables, the $\lambda_n$'s are non-negative real numbers,
and $(\phi_n)_{n \in \mathbb{N}}$ is an orthonormal basis of $L^2(\mu)$
where $\mu$ is a measure on $\mathbb{R}^2$. 
It gives the eigendecomposition of the covariance kernel $k$ (see e.g. \cite{williams2006gaussian}):
$$ k(x,x') = \sum_{n \in \mathbb{N}} \lambda_n \phi_n(x) \phi_n(x').$$
This extends to functions the usual PCA, which diagonalizes the empirical covariance matrix of numerical data.
FPCA now depends on the choice of the functional space $\mathcal{H}$, associated to $k$, called RKHS (\cite{williams2006gaussian}).
In the literature, there are two main ways to define $\mathcal{H}$.
The first way consists in choosing a standard kernel $k$, often linked to the regularity of the global function (here map).
Then a numerical procedure is used to approximate the KL decomposition of $k$ (see e.g. \cite{Mar2015}, \cite{lamboni2011multivariate}).
The second way is to define $\mathcal{H}$ as a finite-dimensional space spanned by a given basis 
$(\psi_n)_{n=1, \dots, N}$ (see e.g. \cite{ramsay2006functional}). 
Frequent choices for this basis are trigonometric functions (Fourier analysis), splines or wavelets.
In this paper, we have chosen this second way, as wavelets (or splines) seem appropriate to model local heterogeneity of flood maps. 
Then,} it can be shown that doing FPCA of $f_1, \dots, f_n$ on $\mathcal{H}$ is equivalent to doing PCA on the coefficients of $f_1, \dots, f_n$ in $\mathcal{H}$ with the metric given by the Gram matrix of $\psi_1, \dots, \psi_N$, defined by 
$$ G = \left(\int \psi_n(x) \psi_{n'}(x)d\mu(x) \right)_{1 \leq n, n' \leq N}.$$
This metric quantifies the lack of orthogonality of basis functions. 
In particular, for Fourier basis functions and wavelets (but not for splines), which are orthonormal, $G = I_N$ and FPCA comes down to a standard PCA on coefficients.
\review{When the basis is not orthogonal, one may prefer orthonormalizing it, and doing FPCA with the new basis. This is actually equivalent to FPCA on the original basis, as stated in Property~\ref{prop:FPCA_ortho} below. This property can be immediately extended to two basis of $\mathcal{H}$, showing that the result of FPCA does not depend on the chosen basis (orthonormal or not), but only on the (finite-dimensional) space $\mathcal{H}$ that they generate.
\begin{property}[FPCA and orthonormalization] \label{prop:FPCA_ortho}
FPCA for a basis $\psi$ is equivalent to FPCA for an orthonormalized basis obtained from $\psi$. In other words, doing PCA of the basis coefficients with the metric given by the Gram matrix $G$ is equivalent to doing PCA of the orthonormalized basis coefficients with the usual identity metric.
\end{property}
\begin{proof}
Denote $\psi = (\psi_1, \dots, \psi_N)^\top$.
An orthonormalized basis obtained from $\psi$ has the form $R^{-1}\psi$ where $R$ is a square root of $G$ i.e. such that $R R^\top = G$ (see e.g. \cite{redd2012comment}, Lemma 1). Then, using the isometry $\Vert c \Vert^2 _G = c^\top G c = \Vert R^\top c \Vert^2$, doing PCA with the metric $G$ on the basis coefficients $c = (c_1, \dots, c_N)^\top$ is equivalent to doing PCA with the identity metric on the transformed coefficients $R^\top c$, which are the coefficients in the orthonormal basis $R^{-1} \psi$.
\end{proof}
}


\subsection{Spatial data approximation}
\label{sec::background::subsec::SpatialApproximation}

In this paper, FPCA implementation needs to determine a functional basis, where approximates spatial data. Maps can contain local specific behavior as for coastal flooding maps: sharp irregularities in cities explained by the presence of infrastructures, non-flooded areas, etc. Basis systems exist to represent such data, by analysing maps area by area. Among FDA and image processing techniques, B-splines and wavelet basis are commonly used.\\ 



\subsubsection{B-splines basis}
\label{sec::background::subsec::SpatialApproximation::Bsplines}
\begin{figure}[H]
    \centering
    \includegraphics[width=0.7\textwidth]{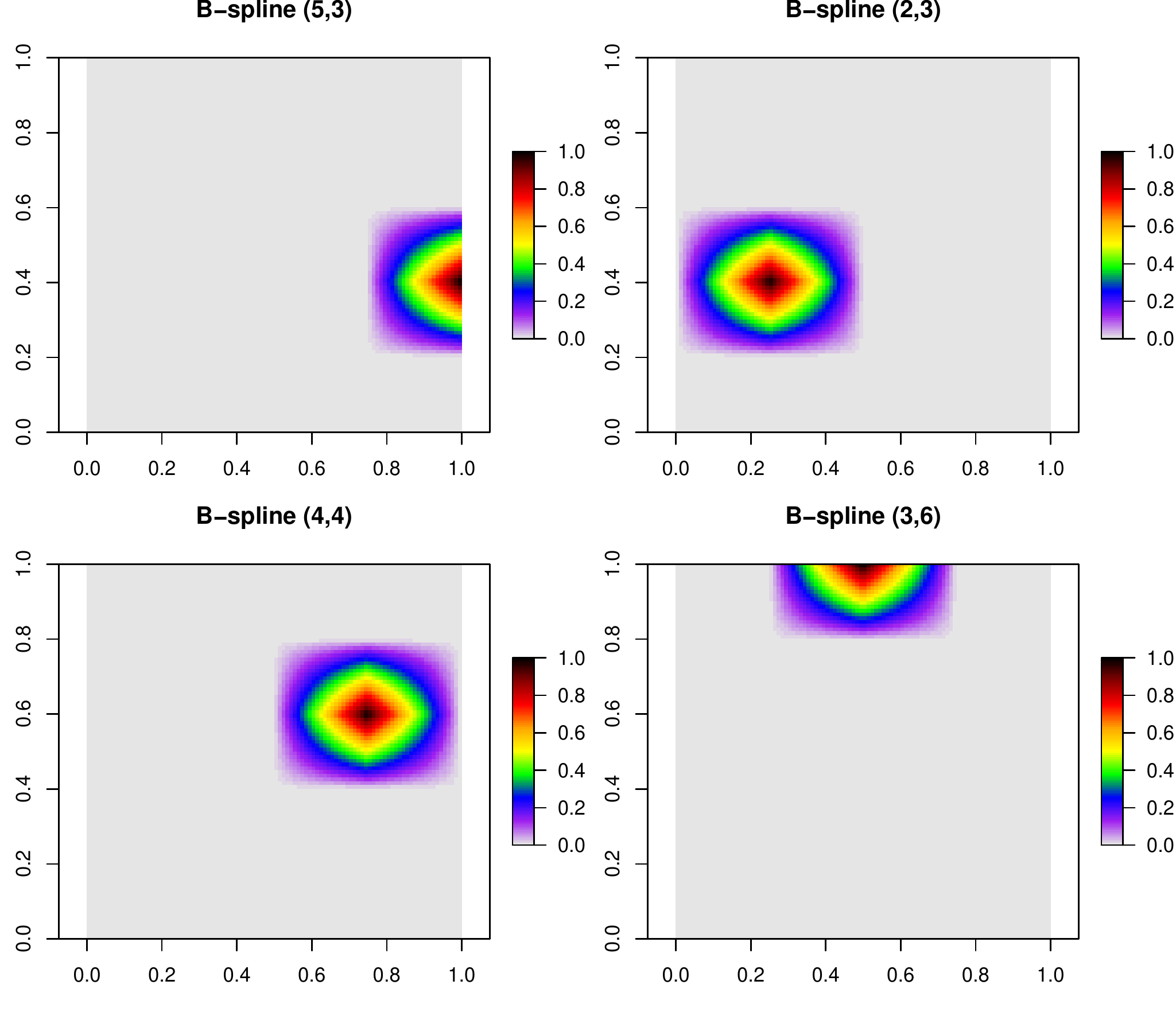}
    \caption{\review{Examples of functions from a two dimensional B-spline basis of degree 1, defined on $[0,1]^2$. The subdivision of the $x$-axis is $\left\lbrace 0, \frac{1}{4}, \frac{1}{2}, \frac{3}{4}, 1 \right\rbrace$. The one of the $y$-axis is $\left\lbrace 0, \frac{1}{5}, \frac{2}{5}, \frac{3}{5},\frac{4}{5},1 \right\rbrace$}}
    \label{fig:ExpleBsplines}
\end{figure}

Splines are piecewise functions defined by polynomials. They are commonly used to approximate non-periodic functional data. Basis systems have been developed for spline functions. In this paper, as the flood maps can be irregular, we consider B-splines basis of degree 1 \cite{ramsay2006functional}, which define a basis for piecewise linear functions. 
They are illustrated on Figure \ref{fig:ExpleBsplines}.
For spatial data, two-dimensional splines can be obtained by tensorisation. More precisely, let two B-splines basis defined on $[0,1]$, denoted $\phi^{(i)}(z_i)=(\phi_1^{(i)}(z_i),\dots,\phi_{K_i}^{(i)}(z_i))^\top$, where $i$ is the coordinate number ($i\in\left\{1,2\right\}$), and $K_i$ is the number of knots per coordinate. \review{We denote $K = K_1K_2$ the number of basis functions}. Then, two-dimensional B-splines are obtained by: 
\review{$$\phi_{k_1, k_2}(z_1, z_2)=\phi_{k_1}^{(1)}(z_1)\phi_{k_2}^{(2)}(z_2), \textnormal{ with }1\leq k_i \leq K_i, \, i = 1, 2.$$}

\subsubsection{Wavelet basis}
\label{sec::background::subsec::SpatialApproximation::wavelet}

\label{sec:background:subsec:wavelet}
\begin{figure}[H]
     \centering
     \includegraphics[width=1\textwidth]{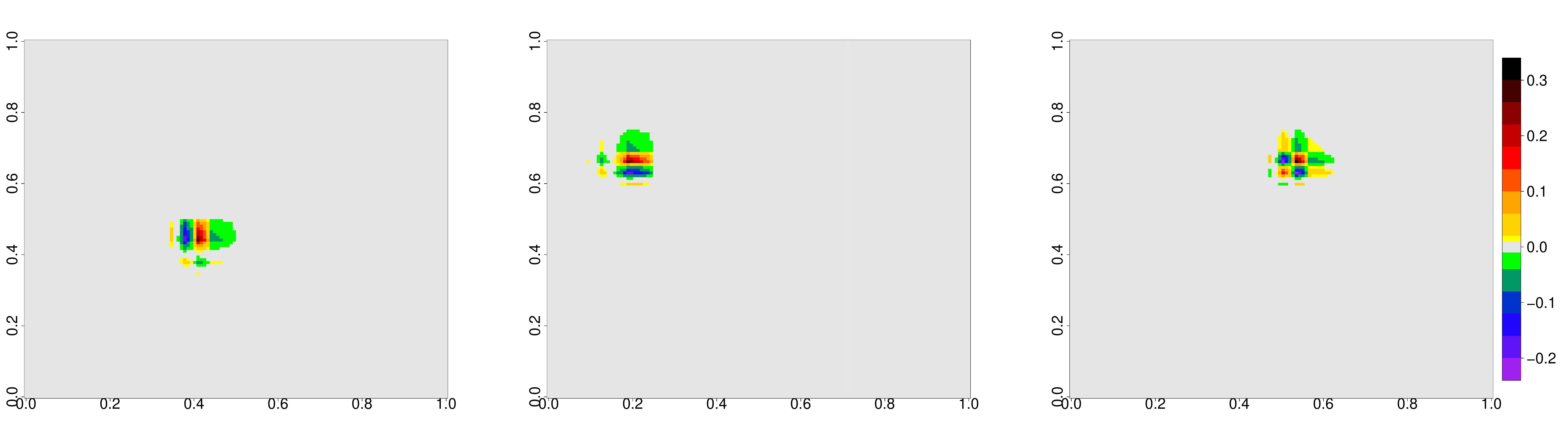}
     \includegraphics[width=1\textwidth]{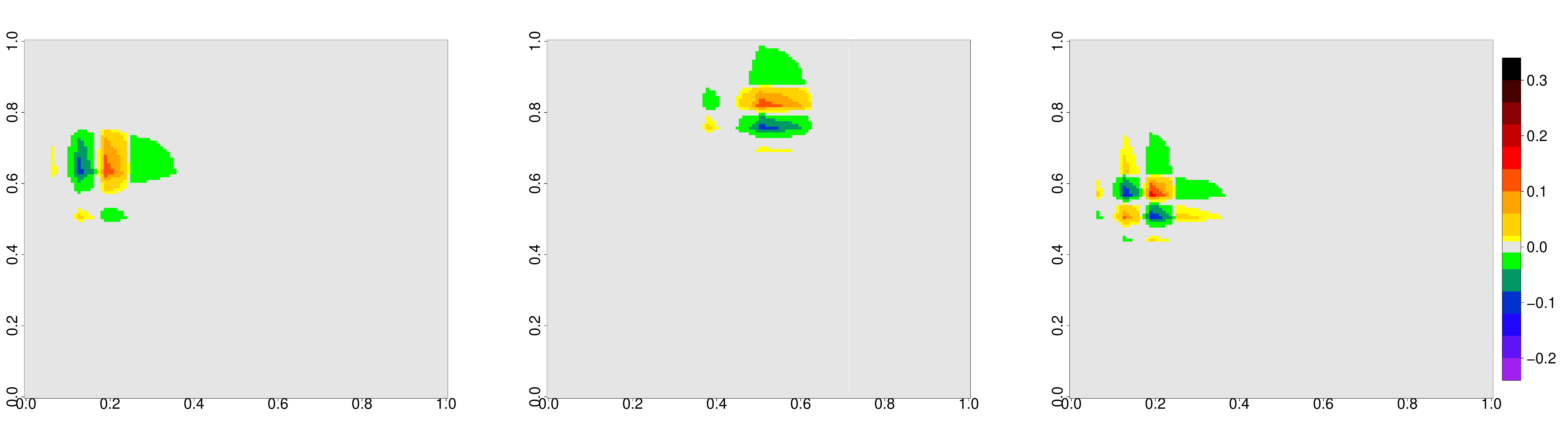}
     \caption{Examples of D4 Daubechies wavelets on \review{$[0,1]^2$, which is 
discretized into a grid of size $128\times 128$. From left to right, the figures represent examples of horizontal, vertical and diagonal wavelets. The top figures are wavelets at scale 3. The bottom figures are wavelets at scale 4.}}
     \label{fig:D4basis}
 \end{figure}

Wavelets $\psi$ are oscillating functions defined on a compact set (i.e. the oscillation exists in a finite duration). They are zero-mean square-integrable functions. Different types of wavelet exist, which is a key strength of wavelet analysis. Daubechies wavelets are widely used in image processing. In this paper, D4 Daubechies wavelets are 
chosen, in order to reach a good tradeoff 
between the size of the support and the selectivity in the frequency domain. Indeed, approximating coastal flooding maps needs a short support due to the local sharp irregularities\footnote{Different wavelets have been tested like Haar wavelets. Best results have been obtained with D4 Daubechies wavelets.}. 
Wavelet basis is built by using translated and dilated versions of a ``mother'' wavelet. Main idea behinds wavelets is to analyse a signal (or image, or a map) according to multiple scales (or resolutions) \cite{mallat1999wavelet}. 
Let us notice that for a multi-resolution analysis, we need at a certain scale to complete the analysis provided by wavelets, with a set of functions which are translated and dilated versions of the ``scaling'' function, associated to the mother wavelet. At a given scale, the coefficients associated with the scaling function are computed with a low-pass filter whereas those obtained with the mother wavelet are computed with a band-pass filter. Examples of D4 Daubechies \cite{daubechies1988orthonormal} wavelets at different scales and translations are illustrated in Figure \ref{fig:D4basis}. For spatial data, as for B-splines, two-dimensional wavelets are obtained by tensorisation.

\subsection{Global sensitivity analysis (GSA)}
\label{sec:background:subsec:GSA}
Global Sensitivity Analysis (GSA) is the set of methods which determine the influence of the input parameters on model output. In this section, we consider models with univariate output as in (\ref{GPfunction}). The most common approach is to use sensitivity indices using ANOVA (Analysis of variance) based on variance decomposition. We consider a vector $\mathbf{X} = (X_1, \dots, X_d)$ 
of independent real random variables, with probability distributions $\mu_1, \dots, \mu_d$. 
We assume that
\begin{equation}
Y=f(\mathbf{X})
\end{equation}is a square-integrable function. Then we can decompose $f(\mathbf{X})$ as a sum of terms of increasing complexity (see e.g. \cite{sobol1993sensitivity}):
\begin{equation}
f(\mathbf{X}) = \underset{\omega\in S}{\sum} f_\omega(\mathbf{X}_\omega)
\label{ANOVA}
\end{equation}where $S=\mathcal{P}(\left\{1,\dots,d\right\})$ is the set of subsets of $\left\{1,\dots,d\right\}$ and $\mathbf{X}_\omega = \left\{(\mathbf{X_l})_{l\in\omega}, \omega\in S\right\}$ is the vector of input variables whose indices are in $\omega\in S$. \review{For $\omega=\emptyset$, we have $f_\emptyset(\mathbf{X}_\emptyset)=f_0$, which is a constant term.}
The decomposition is unique provided that for all set $\omega$ and all strict subset $\omega' \subsetneq \omega$, $\E(f_\omega(\mathbf{X}_\omega) \vert \mathbf{X}_{\omega'}) =0$ holds (with the convention $\E(. \vert X_\emptyset) = \E(.)$). 
Then, the terms of (\ref{ANOVA}) are uncorrelated (orthogonal). Consequently,
the variance of $Y$ can be decomposed:
\begin{equation}
\Var(Y) = \underset{\omega\in S}{\sum} V_\omega
\label{VarDecompo}
\end{equation}where for all $\omega\in S$, $V_\omega = \Var(f_\omega(\mathbf{X}_\omega)) = \int [f_\omega(\mathbf{x}_\omega)]^2 d\mu(\mathbf{x})$. 
Each $f_\omega(\mathbf{X}_\omega)$ is found recursively by conditional expectation on $Y$ knowing $\mathbf{X}_\omega$.
By denoting $d\mu_{-\omega} = \prod_{i,i \notin \omega} d\mu_i(x_i)$, we have for all $i,j$:
\begin{eqnarray*}
f_0 &=& \E(f(\mathbf{X})) = \int f(\mathbf{x})d\mu(\mathbf{x}), \\ f_i(x_i) &=& \E(f(\mathbf{X}) \vert X_i = x_i) - f_0  = \int f(\mathbf{x}) d\mu_{-i}(\mathbf{x}) - f_0, \\
f_{i,j}(x_i,x_j) &=& \E(f(\mathbf{X}) \vert X_i = x_i, X_j = x_j) - f_i\review{(x_i)} - f_j\review{(x_j)} - f_0 \\
&=& \int f(\mathbf{x}) d\mu_{-\{i,j\}}(\mathbf{x}) - f_i\review{(x_i)} - f_j\review{(x_j)} - f_0,
\end{eqnarray*}and more generally, for all $\omega \in S$:
\begin{equation}
f_{\omega}(x_\omega) = \E(f(\mathbf{X}) \vert \mathbf{X}_\omega = \mathbf{x}_\omega) - \sum_{\omega' \subsetneq \omega} f_{\omega'}\review{(\mathbf{x_{\omega'}})} \\
= \int f(\mathbf{x}) d\mu_{-\omega}(\mathbf{x}) - \sum_{\omega' \subsetneq \omega} f_{\omega'}\review{(\mathbf{x_{\omega'}})}.
\end{equation}Equations (\ref{ANOVA}) and (\ref{VarDecompo}) show that when the inputs are independent, 
the variance is decomposed as the sum of contributions of individual effects, second order interactions, and higher order interactions.
Therefore, to quantify the proportion of variance explained by variables in $\omega$, we can use the so-called Sobol indices,
\begin{equation}
\label{Sobol}
SI_\omega = \frac{V_\omega}{\Var(Y)}.
\end{equation}The Sobol indices satisfy $\sum_{\omega\in S} SI_\omega = 1$. 
\cite{homma1996importance} introduces an index to measure the \textit{total effect} of an input parameter: its individual effect and all its interactions with other input variables. Particular interest is given to the first-order indices $SI_i$ ($i \in \left\{1,\dots,d\right\} $) and the first-order total sensitivity indices ($TI_i$):
\begin{equation}
SI_i = \frac{V_i}{\Var(Y)}
\end{equation}and
\begin{equation}
TI_i = \underset{i\in \omega}{\underset{\omega\in S}{\sum}}SI_\omega = 1 - SI_{\left\{1,\dots,d\right\}\backslash \left\{i\right\}}.
\end{equation}

\subsection{Generalized GSA for model with spatial output}
\label{sec:background:subsec:GSI}
In this section, the following simulator is considered:
\begin{equation}
\begin{array}{ccccc}
f &:& \Omega \subseteq \mathbb{R}^d &\rightarrow & \mathbb{L}^2(\mathcal{Z}) \\
  & &            \mathbf{x}                &  \mapsto  & y_\mathbf{x}(\mathbf{z})
\end{array}
\label{model}
\end{equation}where $\mathbf{x}$ is the input vector, $\mathcal{Z}$ is the spatial domain, and $y_\mathbf{x}(\mathbf{z})$ is the output map value at the location $\mathbf{z}$.
For sensitivity analysis, Sobol indices can be computed pointwise as in \cite{marrel2010global}, for each location $\mathbf{z}$. However, it is interesting to assess the global spatial influence of the inputs over the domain. To that end, 
\cite{lamboni2011multivariate}
have introduced so-called 
generalized sensitivity indices (GSI), for multivariate outputs. \cite{gamboa2014sensitivity} have added theoretical arguments to confirm their definition.

\begin{mydef}
The generalized sensitivity index of $y_\mathbf{X}(\mathbf{z})$ with respect to $\mathbf{x}_\omega$ $(\omega\subseteq \left\{1,\dots,d \right\})$, is:
\begin{equation}
GSI_\omega = \frac{\textnormal{Trace}\left(\mathbb{C}\textnormal{ov}\left(\mathbb{E}_{\mathbf{X}}[y_\mathbf{X}(\mathbf{z})| \mathbf{X}_\omega]\right)\right)}{\textnormal{Trace}\left(\mathbb{C}\textnormal{ov}\left(y_\mathbf{X}(\mathbf{z})\right)\right)}
\label{eq:GSI}
\end{equation}
with $\textnormal{Trace}\left(\mathbb{C}\textnormal{ov}\left(y_\mathbf{X}(\mathbf{z})\right)\right) = \int_\mathcal{Z} \mathbb{V}\textnormal{ar}\left(y_\mathbf{X}(\mathbf{z})\right)d\mu(\mathbf{z})$
(and similar definition for the numerator).
The generalised total sensitivity index w.r.t. $\mathbf{X}_j$ is defined by $GTSI_{j} = \sum_{\omega,j\in\omega}GSI_\omega$.
\label{def:GSI}
\end{mydef}

The model output of (\ref{model}) is infinite dimensional, which makes difficult a direct analysis. Therefore, there is a need to reduce dimension.
\cite{lamboni2011multivariate} proposed to use principal component analysis (PCA), 
after discretizing $\mathcal{Z}$. 
Then, the following decomposition is obtained:
\begin{equation}
y_\mathbf{x}(\mathbf{z}) = \mu(\mathbf{z}) + \sum_{k=1}^K \theta_k\xi_k(\mathbf{z}),\ \forall\mathbf{x}\in \Omega
\end{equation}with $\mu(\mathbf{z}) = \mathbb{E}(Y_\mathbf{X}(\mathbf{z}))$, and $(\theta_k)_{k=1,\dots,k}$ are the coordinates of $y_\mathbf{x}(\mathbf{z})$ on the eigenvectors basis $(\xi_k(\mathbf{z}))_{k=1,\dots,K}$. Then, $GSI$ is computed thank to Property \ref{prop:GSI_PCA}.

\begin{property}[\cite{lamboni2011multivariate}]
For all $\omega\subseteq \left\{1,\dots,d \right\}$, the generalized sensitivity index satisfy:
$$GSI_\omega = \frac{\sum_{k}^K \lambda_k SI_{\omega,k}}{\sum_{k=1}^K \lambda_k}$$
where $\lambda_k$ is the $k^\textrm{th}$ eigenvalue, $SI_{\omega,k}$ is the Sobol index on the $k^\textrm{th}$ principal component, which corresponds to the influence of $\mathbf{x}_\omega$ on $\theta_k$ value. Furthermore, $0<GSI_\omega<1$ and $\sum_{\omega\subseteq \{1, \dots, d\}}GSI_\omega=1$. 
\label{prop:GSI_PCA}
\end{property}

\subsection{Gaussian process regression}
\label{sec:background:subsec:GP}
The methodology introduced in this paper can be used for any kind of metamodels. For the application, we focus on one of the most famous, the Gaussian process (GP) regression (also called Kriging).
It is used to interpolate the coordinates of the principal components.
We provide here a brief presentation for the case of a single scalar output. More details can be found in \cite{williams2006gaussian}.

Let 
$f:\mathcal{X}\subseteq \mathbb{R}^d \rightarrow \mathbb{R}
\label{GPfunction}$
be a multivariate function representing the simulator.
Consider a learning set, or design of experiments, $\mathbf{x}^{(1)},\dots,\mathbf{x}^{(n)}$ and associated observations $y_i=f(\mathbf{x}^{(i)})$ ($i=1,\dots,n$). In the probabilistic interpretation of Kriging, the function $f$ is seen as a realization of a Gaussian process $Y(\mathbf{x})$ of mean $m(\mathbf{x})$ and covariance function, or kernel, $C(\mathbf{x},\mathbf{x}')$.
The kernel contains the spatial dependencies between $\mathbf{x}$ and $\mathbf{x}'$. Under stationary assumption, the kernel depends only on $x-x'$, and is often chosen as a decreasing function of the distance $\vert x - x'\vert$. In this paper, we have used the tensor product Mat\'ern $5/2$ kernel, which is a standard choice.

The prediction at a new input $\mathbf{x}^*$ is obtained as the conditional probability distribution of $Y(\mathbf{x}^*)$ knowing $Y(\mathbf{x}^{(i)})=y_i$ ($i=1,\dots,n$). By properties of Gaussian vectors, one obtains closed-form expression for its mean $\hat{y}(\mathbf{x}^*)$ and its variance $\sigma^2_y(\mathbf{x}^*)$:
\begin{equation}
\label{eq:GP}
\begin{array}{ccc}
\hat{y}(\mathbf{x}^*) &= & m(\mathbf{x}^*) + c(\mathbf{x}^*)^\top \mathbf{C}^{-1}y \\
\sigma^2_y(\mathbf{x}^*)&=& C(\mathbf{x}^*,\mathbf{x}^*) - c(\mathbf{x}^*)^\top \mathbf{C}^{-1}c(\mathbf{x}^*)
\end{array}
\end{equation}  
where $\mathbf{C}=(C(\mathbf{x}^{(i)},\mathbf{x}^{(j)}))_{1\leq i,j\leq n}$ is the covariance matrix at design points, 
and $c(\mathbf{x}^*) = \review{(}C(\mathbf{x}^{*},\mathbf{x}^{(i)}))_{1\leq i\leq n}$ is the vector of covariances between the new point and design points. Notice that by construction the prediction is an interpolator: $\hat{y}(\mathbf{x}^{(i)}) = y_i$. In practice, the kernel parameters are estimated, e.g. by maximum likelihood, and other expressions for the conditional mean and variance, known as universal Kriging formula, can be derived. They account for the additional uncertainty coming from estimation error (see e.g. \cite{williams2006gaussian}).


%
%
\section{Metamodels for spatial outputs based on FPCA}
\label{sec:metamodelsFPCA}
\subsection{\review{Methodology with an orthonormal basis}}

We consider the simulator as defined in (\ref{model}). We assume that we know $n$ simulations of $f$: $\left\{(\mathbf{x}_i,y_\mathbf{x_i}(\mathbf{z})),i=1,\dots,n\right\}$. We aim at predicting the map $f(\mathbf{x}^*)$ for a new point $\mathbf{x}^*$.

Each map is seen as a function in $\mathbb{L}^2(\mathcal{Z})$. In practice, it is necessary to go down to finite dimensions. Instead of discretizing the maps in space by using a finite number of locations, we consider a functional subspace of finite dimension by using basis functions, denoted by $\Phi(\mathbf{z})=\left(\phi_1(\mathbf{z}),\dots,\phi_K(\mathbf{z})\right)^\top$. \review{$K$ is the number of basis functions}. For all $\mathbf{x}\in\Omega$, we then have:
\begin{equation}
\label{PhiDecompo}
 y_\mathbf{x}(\mathbf{z}) = \sum_{k=1}^K \alpha_k(\mathbf{x})\phi_k(\mathbf{z})=\alpha(\mathbf{x})^\top\Phi(\mathbf{z})
\end{equation}where $\alpha(\mathbf{x}) = (\alpha_1(\mathbf{x}),\dots,\alpha_K(\mathbf{x}))$ is a vector of coefficients. 
Then, predicting the spatial map $ y_\mathbf{x^\star}(\mathbf{z})$ at a new point $\mathbf{x}^\star$ comes down to predict the $K$ real numbers $\alpha_1(\mathbf{x^\star}),\dots,\alpha_K(\mathbf{x^\star})$ at $\mathbf{x^\star}$.\\

\review{$K$ should be chosen in order to minimize the root mean square error of the approximation of $y_\mathbf{x}(\mathbf{z})$ on the $\Phi(\mathbf{z})$ basis.} However, in order to approximate accurately the spatial maps, the size of the basis function $K$ should be large \emph{a priori}. 
To further reduce dimension, two procedures are applied sequentially: selection of coefficients and PCA on the selected coefficients.\\

We detail the selection step, for which a careful treatment is necessary. We \review{first} assume that $\Phi(\mathbf{z})$ is an orthonormal basis \review{, but an alternative based on sparse regression is also presented in the next section ``Variants, without orthonormalization''}. An orthonormalization method may be applied first, such as Gram-Schmidt \cite{bjorck1994numerics} or a specific procedure developed for B-splines \cite{qin2000general}, \cite{redd2012comment} \cite{liuob}. \review{
We notice that the energy can be decomposed as follows:}

\begin{equation}
||y_\mathbf{x}||_2^2 = \int y_\mathbf{x}(\mathbf{z})^2 d\mu(\mathbf{z}) = \sum_{k=1}^K \alpha_k(\mathbf{x})^2.
\end{equation}

\review{Therefore, each coefficient $\alpha_k(\mathbf{x})$, $k=1,\dots,K$, corresponds to a part of the energy. The importance of a coefficient $\alpha_k(\mathbf{x})$ can be quantified by the ratio} $\frac{\alpha_k(\mathbf{x})^2}{\sum_{k'=1}^K \alpha_{k'}(\mathbf{x})^2 }$. However, such ratios depend on $\mathbf{x}$, which is an issue for prediction on a new point $\mathbf{x}^*$. Hence, we consider instead the mean proportion of energy:\\

\begin{equation}
\label{lambda}
    \lambda_k = \mathbb{E}\left[\frac{\alpha_k(\mathbf{X})^2}{\sum_{k'=1}^K \alpha_{k'}(\mathbf{X})^2 }\right]
\end{equation}

\review{ $\lambda_k$ does not depend on $\mathbf{x}$. In practice, we approximate the expectation by the empirical mean on the learning set $(\mathbf{x}^{(i)})_{i =1,\dots,n}$, which is a good approximation if these design points have been drawn from $\mu$  in $\Omega$ (e.g. a space-filling design if $\mu$ is the uniform distribution). 
Now, the $\lambda_k$'s can be sorted in decreasing order. 
We denote by $(k)$, $k=1,\dots,K$, the corresponding indices. 
We truncate the basis by selecting the $\tilde{K}$ ($\leq K$) largest coefficients such that 
$\sum_{k=1}^{\tilde{K}}\lambda_{(k)}\leq p$, where $p\in [0,1]$ is the total mean proportion of energy. 
For the other indices $(\tilde{K}+1),\dots (K)$, thus corresponding to the smallest contributions in energy, 
the coefficients $\alpha_{(k)}(\mathbf{x})$ are made constant and replaced by their empirical mean: $\hat{\alpha}_{(k)}(\mathbf{x})= \frac{1}{n}\sum_{i=1}^n\alpha_{(k)}(\mathbf{x}^{(i)})$.
Notice that the complexity of this selection step is $O(nK)$,
including the computations of the $\lambda_k$'s and their ranking.
This is moderate and negligible compared to the complexity of the next PCA step, in $O(\min(n, \tilde{K})^3$).}\\


Then, after this selection step, 
we apply a standard PCA \review{on the $\tilde{K}$ selected coefficients. We denote $n_{PC}$ the number of principal components which are modeled.} Then, we predict each coordinate \review{on the $n_{PC}$ first principal components} by separate GPs, which provides a prediction for \review{$(\alpha_{(k)}(\mathbf{x}^\star))_{k=1,\dots,\tilde{K}}$. 
One may think of using multivariate GPs (such as co-kriging models).
However, in addition to increasing the computational burden, 
its benefits compared to separate GPs may be limited as principal components are uncorrelated.
}\\


The whole methodology is summarized in Algorithm \ref{algo:metamodelsFPCA}. In practice, the parameters $p$ and $n_{PC}$ are tuned by cross-validation. \review{We note that the size of the truncation $\tilde{K}$ can directly be tuned instead of $p$.} This is detailed in the application part of the paper.

%
%

\begin{algorithm}[H]
\caption{Aim: To predict $f(\mathbf{x}^\star)=y_{\mathbf{x}^\star}(\mathbf{z})$, $\mathbf{z}\in\mathcal{Z}$}
\label{algo:metamodelsFPCA}
\SetAlgoLined

\KwIn{
$\left\{(\mathbf{x}_i,y_\mathbf{x_i}(\mathbf{z})),i=1,\dots,n\right\}$; $\Phi(\mathbf{z})=\left(\phi_1(\mathbf{z}),\dots,\phi_K(\mathbf{z})\right)^\top$ (a functional basis); $p$ (mean proportion of energy); $n_{PC}$ (number of principal components)
}
\bigskip
\KwOut{$\hat{f}(\mathbf{x}^\star)=\hat{y}_{\mathbf{x}^\star}(\mathbf{z})$\;}
 \bigskip

\begin{enumerate}
    \item If $\Phi(\mathbf{z})$ is not an orthonormal basis, orthonormalize it with a suitable method. 
    For simplicity, the new basis is still denoted $\Phi(\mathbf{z})$.
    
    \item Decompose the $(y_{\mathbf{x}^{(i)}}(\mathbf{z}))_{i=1,\dots,n}$ in the $\Phi(\mathbf{z})$ basis. 
    
 \item Sort the basis coefficients $(\alpha_k(\mathbf{x}))_{k=1,\dots,K}$ in the decreasing order of the criterion (\ref{lambda}). We denote $(k)$, $k=1,\dots,K$, the indices of the coefficients following the new order.
    Select the $\tilde{K}\ll K$ most important coefficients such as $\sum_{k=1}^{\tilde{K}} \lambda_{(k)}\leq p$ (Equation (\ref{lambda})).
    
 \item Apply PCA in $\mathbb{R}^{\tilde{K}}$ on the dataset of  coefficients evaluated at design points $(\alpha(\mathbf{x}^{(i)}))_{i = 1, \dots, n}$. Choose the first $n_{PC}$ principal components. Denote $t_1(\mathbf{x}^{(i)}),\dots, t_{n_{PC}}(\mathbf{x}^{(i)})$, $i=1,\dots,n$, the coordinates on the first principal components, and $w_1,\dots,w_{n_{PC}}$, the associated eigenvectors.
    
    \item For each principal component $l=1,\dots,n_{PC}$, predict $t_l(\mathbf{x}^\star)$ (denoted $\hat{t}_l(\mathbf{x}^\star)$) by GP regression (see (\ref{eq:GP})), based on the observation $t_l\mathbf{x}^{(i)})$ ($i=1,\dots,n$). 
         
    \item Predict the coefficients $\alpha_k(\mathbf{x}^\star)$. 
    \begin{itemize}
        \item for $k=\tilde{K}+1,\dots,K$, $\hat{\alpha}_{(k)}(\mathbf{x}^\star) = \frac{1}{n}\sum_{i=1}^n \alpha_{(k)}(\mathbf{x}^{(i)}) $.
        \item for $k=1,\dots,\tilde{K}$, predict the coefficients by their coordinates estimation on the principal components: $\hat{\alpha}_{(k)}(\mathbf{x}^\star) = \sum_{l=1}^{n_{PC}} \hat{t}_l(\mathbf{x}^\star)w_l$ 
    \end{itemize}
    
    \item Compute the prediction $y_{\mathbf{x}^\star}(\mathbf{z})$ with the predicted coefficients $\hat{\alpha}_k({x}^\star)$ from Equation (\ref{PhiDecompo}).
\end{enumerate}

\end{algorithm}

\review{
\subsection{Variants, without orthonormalization.}
The steps 1-3 of Algorithm~\ref{algo:metamodelsFPCA} 
describes how to select the basis coefficients with a $L^2$ decomposition, which has a physical meaning in terms of energy.
Alternatively, one can think of applying statistical sparse techniques, based on $L^1$ penalty.
For instance, the Lasso regression \cite{tibshirani1996regression} is written as follows.
For a given $\mathbf{x} \in \mathcal{X}$, let $\lambda(\mathbf{x}) > 0$ be a penalty parameter.
Without loss of generality, we assume that $y_\mathbf{\mathbf{x}}(\mathbf{z})$ has been centered with respect to $\mathbf{z}$. 
Then, the coefficients $\alpha_k(\mathbf{x})$ are estimated by solving the penalized regression problem
\begin{equation}
\label{eq::Lasso}
    \underset{\alpha_1(\mathbf{x}), \dots, \alpha_K(\mathbf{x})}{\min} \Vert y_\mathbf{x}(\mathbf{z}) - \sum_{k = 1}^K \alpha_k(\mathbf{x}) \phi_k(\mathbf{z}) \Vert^2
    + \lambda(\mathbf{x}) \sum_{k=1}^K \vert \alpha_k(\mathbf{x}) \vert.
\end{equation}
In practice the $L^2$ norm is replaced by its discretization on $\mathbf{z}$.
As it is well known, this optimization problem induces sparsity, 
and forces some coefficients $\alpha_k(\mathbf{x})$ to be equal to zero. 
However, as for the selection based on energy, this selection depends on $\mathbf{x}$.
Hence, we can consider some global criterion such as $\E(\alpha_k(\mathbf{x}))$ or $\mathbb{P}(\alpha_k(X) \neq 0)$, 
to make the choice of the $k$ values independent of $\mathbf{x}$. 
Thus, although the Lasso technique induces sparsity for a single map $y_{\mathbf{x}}(.)$, 
this is not true for the collection of maps (when $\mathbf{x}$ varies), 
and we must also specify the desired number $\tilde{K}$ of selected coefficients. 
The rest of the algorithm (steps 4-7) is unchanged.\\
Clearly, one strength of that selection variant is that it can be applied to any functional basis, 
without the need of orthonormalization.
On the other hand, it adds a tuning parameter $\lambda(\mathbf{x})$ for all $\mathbf{x} \in \mathcal{X}$,
which increases the global computational cost. 
A less expensive alternative is to consider a \emph{common} penalty parameter $\lambda$ for all $\mathbf{x} \in \mathcal{X}$,
assuming that the outputs $y_\mathbf{x}(.) $ have a similar level of regularity when $\mathbf{x}$ varies.
Then $\lambda$ could be tuned by cross-validation, in addition to $\tilde{K}$, in Algorithm~\ref{algo:metamodelsFPCA}.
}

%
%

\section{Extension of generalized sensitivity indices} 
\label{sec:GSIprop}

In this section, the simulator (Eq. \ref{model}) is considered. In section \ref{sec:background:subsec:GSI}, generalized sensitivity indices ($GSI$) have been defined for such model (cf. Definition \ref{def:GSI}). By using Property \ref{prop:GSI_PCA}, they can be computed using PCA, which is allowed due to the orthonormality of the eigen vectors (or functions) basis. Here, we extend this property to any kind of functional basis. 

\begin{property}
Let  $\phi_1, \dots, \phi_K$ be a set of functions with Gram matrix
$\review{\mathbf{G}}=\int \phi(\mathbf{z})\phi(\mathbf{z})^\top d\mu(\mathbf{z})$. Assume that the spatial output $y_x(z)$ is decomposed as: 

\begin{equation}
y_\mathbf{X}(\mathbf{z}) = \sum_{k=1}^K\alpha_k(\mathbf{X})\phi_k(\mathbf{z}).
\end{equation}Denote by $\alpha(\mathbf{X})= (\alpha_1(\mathbf{X}),\dots,\alpha_K(\mathbf{X}))^\top$the vector of coefficients.
Then the GSI of $y$ is given by:
\begin{equation}
GSI_\omega  =  \frac{\textnormal{Trace}(\mathbb{C}\textnormal{ov}(\mathbb{E}[\alpha(\mathbf{X})|\mathbf{X}_\omega]) \review{\review{\mathbf{G}}})}{\textnormal{Trace}(\mathbb{C}\textnormal{ov}(\alpha(\mathbf{X})) \review{\review{\mathbf{G}}})}
\label{eq:GSIextent}
\end{equation}
\label{prop:GSIextent}

\end{property}

\begin{proof}
Recall that GSI are defined by (see Def. \ref{def:GSI}):
$$
GSI_\omega = \frac{\textnormal{Trace}\left(\mathbb{C}\textnormal{ov}\left(
\mathbb{E}[y_\mathbf{X}(\mathbf{z})| \mathbf{X}_\omega]
\right)\right)}
{\textnormal{Trace}\left(\mathbb{C}\textnormal{ov}\left(
y_\mathbf{X}(\mathbf{z})
\right)\right)}=
\frac{\int_\mathcal{Z} \mathbb{V}\textnormal{ar}\left( \mathbb{E} \left[y_\mathbf{X}(\mathbf{z})|\mathbf{X}_\omega\right]\right) d\mu(\mathbf{z})}{\int_\mathcal{Z} \mathbb{V}\textnormal{ar}\left( y_\mathbf{X}(\mathbf{z})\right) d\mu(\mathbf{z})}.
$$For the denominator, we have:
$$
\begin{array}{lll}
\mathbb{V}\textnormal{ar}(y_\mathbf{X}(\mathbf{z})) 
&=& \sum_{k,l=1}^K \mathbb{C}\textnormal{ov}(\alpha(\mathbf{X}))_{k,l}\phi_k(\mathbf{z})\phi_l(\mathbf{z}).
\end{array}
\label{var}
$$For the numerator, by linearity of conditional expectation, we have:
$$
\begin{array}{lll}
\mathbb{V}\textnormal{ar}(\mathbb{E}[y_\mathbf{X}(\mathbf{z})|\mathbf{X}_\omega]) 
&=& \sum_{k,l=1}^K \mathbb{C}\textnormal{ov}(\mathbb{E}[\alpha(\mathbf{X})|\mathbf{X}_\omega])_{k,l}\phi_k(\mathbf{z})\phi_l(\mathbf{z}).
\end{array}
\label{VarExp}
$$Therefore, we obtain:
$$
GSI_\omega = \frac{\sum_{k,l=1}^K \mathbb{C}\textnormal{ov}(\mathbb{E}[\alpha(\mathbf{X})|\mathbf{X}_\omega])_{k,l}  \int_{\mathcal{Z}}\phi_k(\mathbf{z})\phi_l(\mathbf{z})d\mu(\mathbf{z})}{ \sum_{k,l=1}^K  \mathbb{C}\textnormal{ov}(\alpha(\mathbf{X}))_{k,l} \int_{\mathcal{Z}} \phi_k(\mathbf{z})\phi_l(\mathbf{z}) d\mu(\mathbf{z})}.
$$Finally, with $\review{\review{\mathbf{G}}}=\int \phi(\mathbf{z})\phi(\mathbf{z})^\top d\mu(\mathbf{z})$, 
$$
GSI_\omega = \frac{\sum_{k,l=1}^K \mathbb{C}\textnormal{ov}(\mathbb{E}[\alpha(\mathbf{X})|\mathbf{X}_\omega])_{k,l} \review{G_{k,l}}}{ \sum_{k,l=1}^K  \mathbb{C}\textnormal{ov}(\alpha(\mathbf{X}))_{k,l} \review{G_{k,l}}}= \frac{\textnormal{Trace}(\mathbb{C}\textnormal{ov}(\mathbb{E}[\alpha(\mathbf{X})|\mathbf{X}_\omega]) \review{\review{\mathbf{G}}})}{\textnormal{Trace}(\mathbb{C}\textnormal{ov}(\alpha(\mathbf{X})) \review{\mathbf{G}})}
$$where the last equality comes from the property $ \textnormal{Trace}(AB^\top) = \sum_{k,l}A_{k,l}B_{k,l} $, valid for all matrices $A$ and $B$.
\end{proof}





%
%
\section{An analytical test case}
\label{sec:TestCase}
In this section, GP metamodelling using standard PCA, FPCA based on wavelet basis, FPCA based on B-splines basis (respectively denoted $\textnormal{GP}^{\textnormal{PCA}}$, $\textnormal{GP}^{\textnormal{FPCA}}_{\textnormal{wavelet}}$, and $\textnormal{GP}^{\textnormal{FPCA}}_{\textnormal{B-splines}}$) are applied on \review{an} analytical case (presented in Section \ref{sec:TestCase:context}). Then, section \ref{sec:TestCase:subsec:calibration} explains how wavelet and B-spline basis are defined.
The optimal parametrization of FPCA for metamodelling procedure is selected using a cross-validation method (see section \ref{sec:TestCase:subsec:calibration}). Next, the comparison of all methods is made (Section \ref{sec:TestCase:subsec:prediction}). Finally, generalized sensitivity indices are implemented using simulations obtained by FPCA-based method. All implementations are performed using the statistical programming language
\textbf{R} \cite{ManuelR}. \\


\review{A \textbf{R} package, named \textbf{GpOutput2D} \cite{package}, has been developed. \textbf{GpOutput2D} contains functions for applying FPCA and GP regression modelling on two-dimensional functional data. It is based on other \textbf{R} packages for wavelets and orthonormal B-splines decomposition, and for kriging models: \textbf{waveslim},  \textbf{orthogonalsplinebasis}, \textbf{DiceKriging} and \textbf{kergp}.}\\ 

\review{For GSA, the \textbf{sobolSalt} function from \textbf{sensitivity} package has been used to compute Sobol indices on each principal component.}

\subsection{Description of the Campbell2D function}
\label{sec:TestCase:context}
 The performance
 of $\textnormal{GP}^{\textnormal{PCA}}$, $\textnormal{GP}^{\textnormal{FPCA}}_{\textnormal{wavelet}}$, and $\textnormal{GP}^{\textnormal{FPCA}}_{\textnormal{B-splines}}$ are compared on an analytical test case, used by \cite{marrel2010global}, namely
 the Campbell2D function. This function has eight inputs (d=8) and a spatial map as output (e.g. a function which depends on two inputs ($\mathbf{z}=(z_1,z_2)$) corresponding to spatial coordinates).\\

\begin{equation}
\begin{array}{ccccc}
f &:& [-1,5]^8 &\rightarrow & \mathbb{L}^2([-90,90]^2)\\
  & &  \mathbf{x} = (x_1,\dots,x_8) &\mapsto&  y_\mathbf{x}(\mathbf{z})
\end{array}
\end{equation}

where $\mathbf{z}=(z_1,z_2)\in [-90,90]^2$, $x_j\in [-1,5]$ for $j=1,\dots,8$, and 

\begin{equation}
\label{eq:Campbell2D}
\begin{array}{ll}
y_\mathbf{x}(z_1,z_2) =& x_1 \exp\left[-\frac{(0.8z_1+0.2z_2-10x_2)^2}{60x_1^2}\right] + (x_2+x_4)\exp\left[\frac{(0.5z_1+0.5z_2)x_1}{500}\right] +\\
  &  x_5(x_3 -2)\exp\left[-\frac{(0.4z_1+0.6z_2-20x_6)^2}{40x_5^2}\right] +\\ &(x_6+x_8)\exp\left[\frac{(0.3z_1+0.7z_2)x_7}{250}\right]
\end{array}
\end{equation}

\begin{figure}[H]
\begin{center}
\includegraphics[width=0.32\textwidth]{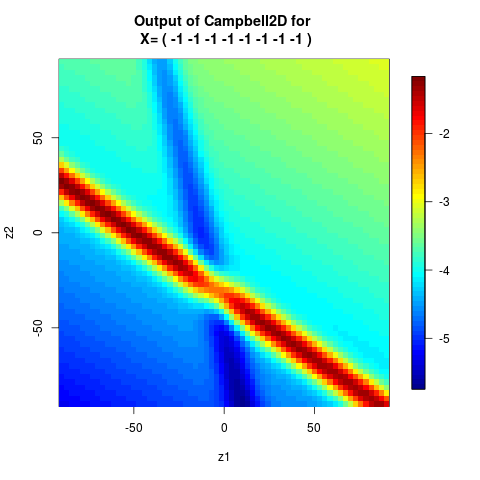}
\includegraphics[width=0.32\textwidth]{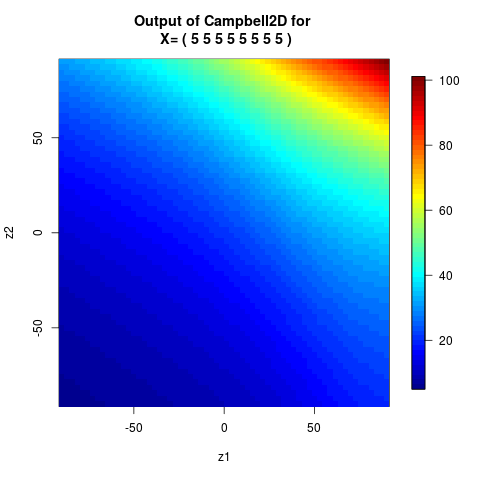}
\includegraphics[width=0.32\textwidth]{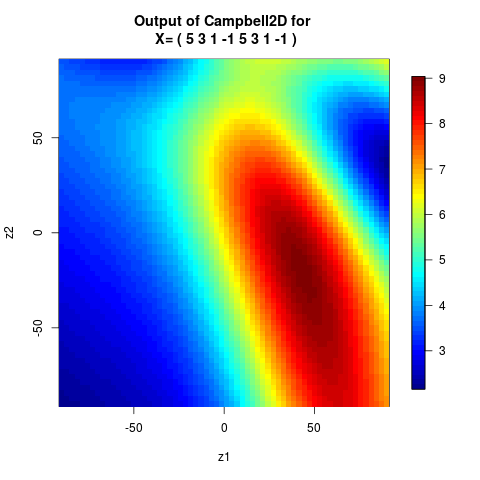}
\caption{Example of Campbell2D spatial outputs. From left to right, $\mathbf{x}=(-1,-1,-1,-1,-1,-1,-1,-1)$, $\mathbf{x}=(5,5,5,5,5,5,5,5)$, and $\mathbf{x}=(5, 3, 1, -1, 5, 3, 1, -1)$.}
\label{Campbell2DOutput}
\end{center}
\end{figure}

Figure \ref{Campbell2DOutput} shows examples of Campbell2D outputs. The output map presents strong spatial heterogeneities, sometimes with sharp boundaries. Furthermore, the spatial distribution is different according to the $\mathbf{x}$ values. A learning sample of size $n=200$ is considered, with a space-filling design of experiment 
constructed using a Latin Hypercube Sampling (LHS)
design optimized by the SA algorithm \cite{dupuy2015dicedesign}, 
(implemented on the \textit{DiceDesign} R package). The design points are denoted $\mathbf{x}^{(i)}$, and the associated output map, $y_{\mathbf{x}^{(i)}}(\mathbf{z})$, $i=1\dots,n$. For the application, the spatial domain $[-90,90]^2$ is discretized on an uniform grid of dimension $64\times 64$. Note that both dimensions must be a power of two, a requirement of wavelet decomposition.\\

\subsection{Choice of FPCA parameters}
\label{sec:TestCase:subsec:calibration}

For wavelet decomposition, D4 Daubechies wavelets are used in this paper. Multiresolution approximation of the output maps needs to define the number of resolutions (also called level of decomposition) \cite{mallat1999wavelet}. For B-splines, splines of degree $1$ are considered, 
and knots are chosen equally spaced. For simplicity, the same number is considered for both dimensions. Wavelet and B-spline basis are selected such that the mean square error between maps of the learning sample and their approximations is minimized.




A
$k$-fold cross-validation \cite{hastie2009cross}, with $k=10$, is used to tune the parameters of $\textnormal{GP}^{\textnormal{FPCA}}_{\textnormal{wavelet}}$ and $\textnormal{GP}^{\textnormal{FPCA}}_{\textnormal{B-splines}}$: number of coefficients ($\review{\tilde{K}}$) for the PCA step of Algorithm \ref{algo:metamodelsFPCA}, number of principal components ($n_{PC}$).  

\begin{figure}[H]
\centering
\includegraphics[width=1\textwidth]{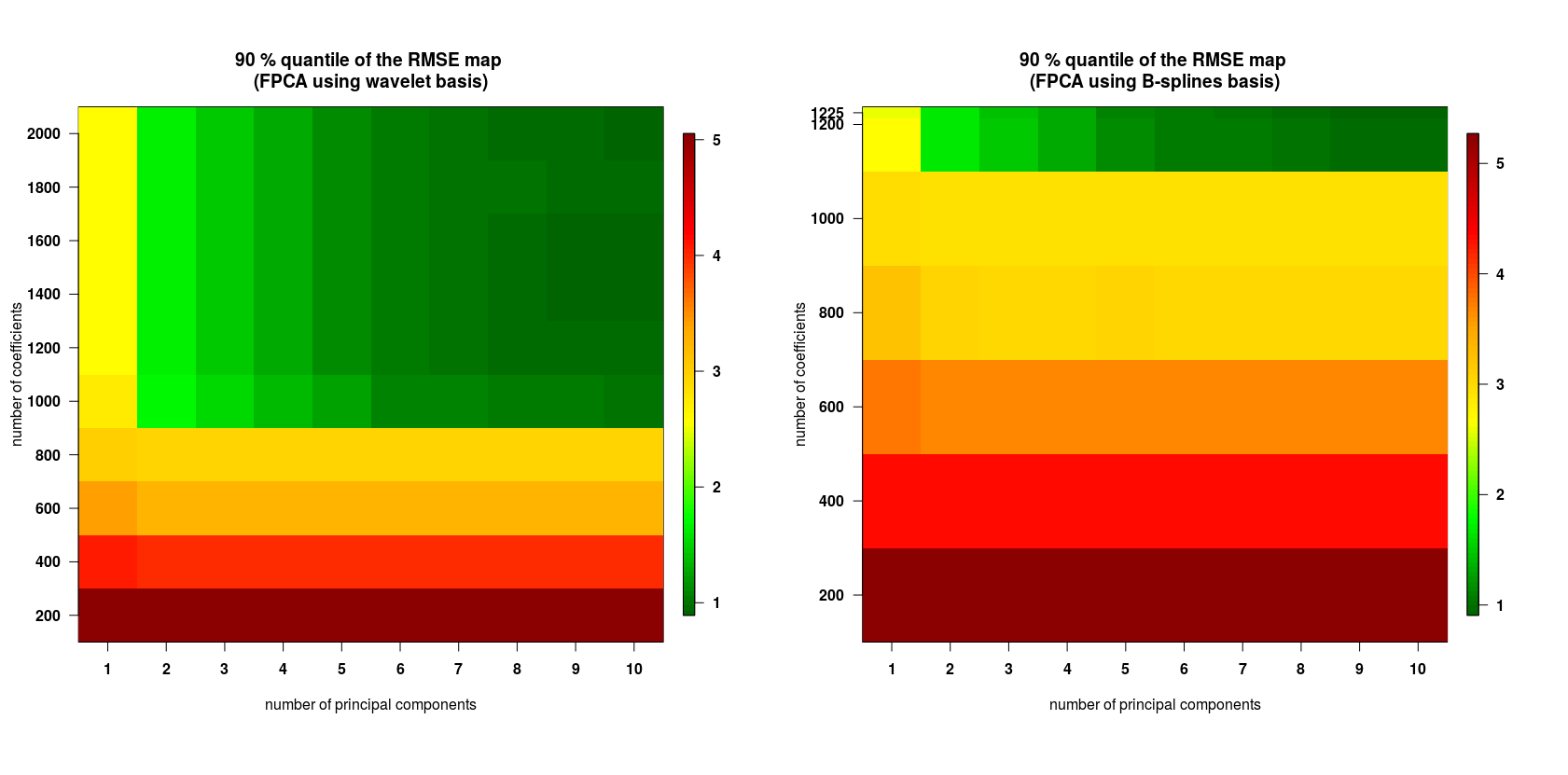}
\caption{$90\%$ quantile of the $10$-fold cross validation RMSE: $\textnormal{GP}^{\textnormal{FPCA}}_{\textnormal{wavelet}}$ (left), $\textnormal{GP}^{\textnormal{FPCA}}_{\textnormal{B-spline}}$ (right).} 
\label{fig:quantile90}
\end{figure}

In order to assess the metamodel predictive performance,
the spatial root mean square error (RMSE) is computed for each sub-sample of the cross-validation procedure as defined in Eq. (\ref{eq:rmsek})
\begin{equation}
\textnormal{RMSE}_{l}(\mathbf{z}) = \sqrt{\frac{1}{n_l}\sum_{i'=1}^{n_l}\left(y_{\mathbf{x}_{i'}^{(l)}}(\mathbf{z}) - \hat{y}_{\mathbf{x}_{i'}^{(l)}}(\mathbf{z})\right)^2}, \ \forall l\in\left\{1,\dots,k\right\}
\label{eq:rmsek}
\end{equation}
%
%
where $(\mathbf{x}_{i'}^{(l)},y_{\mathbf{x}_{i'}^{(l)}}(\mathbf{z}))$ is the $i'$th (input, output) observation of the $l$-th sub-sample of size ${n_l=\frac{n}{k}=20}$, and $ \hat{y}_{\mathbf{x}_{i'}^{(l)}}(\mathbf{z})$ is the estimation of $y_{\mathbf{x}_{i'}^{(l)}}(\mathbf{z})$. Then, a global $k$-fold cross-validation RMSE is computed by averaging the sub-sample RMSEs, as defined in Eq. (\ref{eq:RmseKfold}).
\begin{equation}
\textnormal{RMSE}_{\textnormal{CV}}(\mathbf{z})= \frac{1}{k}\sum_{l=1}^{k} \textnormal{RMSE}_{l}(\mathbf{z})
\label{eq:RmseKfold}
\end{equation}To quantify the local errors, we will use the 90\%-quantile of Eq. \ref{eq:RmseKfold} with respect to $\mathbf{z}$, in order to capture the potentially large spatial variations (compared to the mean of Eq. \ref{eq:RmseKfold}). Figure \ref{fig:quantile90} shows that quantile values according to $\review{\tilde{K}}$ and $n_{PC}$.

For $\textnormal{GP}^{\textnormal{FPCA}}_{\textnormal{wavelet}}$, a convergence of the RMSE is observed from $\review{\tilde{K}}=1200$, for any $n_{PC}$ value. At $\review{\tilde{K}}=1200$, a convergence starts at $n_{PC}=8$. However, lowest values can be seen from $n_{PC}=5$. In average, from the $6$-th principal components, the percentage of explained variance is less than $1\%$. Therefore, to avoid overfitting, $\review{\tilde{K}}=1200$ and $n_{PC}=5$ are considered for $\textnormal{GP}^{\textnormal{FPCA}}_{\textnormal{wavelet}}$. For $\textnormal{GP}^{\textnormal{FPCA}}_{\textnormal{B-splines}}$, at any $n_{PC}$, the RMSE value reaches a minimum value at $\review{\tilde{K}} = 1225 = 35^2$ which corresponds to the overall basis dimension. Although that number is large, it remains reasonable to perform PCA, and is thus chosen for FPCA. Finally, for the same reasons as  $\textnormal{GP}^{\textnormal{FPCA}}_{\textnormal{wavelet}}$, we choose $n_{PC}=5$ principal components. 

\review{For comparison, we use
$n_{PC}=5$ principal components 
for  $\textnormal{GP}^{\textnormal{PCA}}$. The first five principal components correspond to $98\%$ of the total inertia for the three methods. For $\textnormal{GP}^{\textnormal{FPCA}}_{\textnormal{B-splines}}$, all the coefficients are considered for the PCA step of Algorithm \ref{algo:metamodelsFPCA}, which thus represents $100\%$ of the mean energy (spatial variance). For $\textnormal{GP}^{\textnormal{FPCA}}_{\textnormal{wavelet}}$, approximately $29,3\%$ ($\review{\tilde{K}}=1200$) of the wavelet coefficients are kept, which corresponds to almost $100\%$ of the mean energy too.} 

\subsection{Prediction accuracy}
\label{sec:TestCase:subsec:prediction}

\begin{figure}[H]
    \centering
    \includegraphics[width=\textwidth]{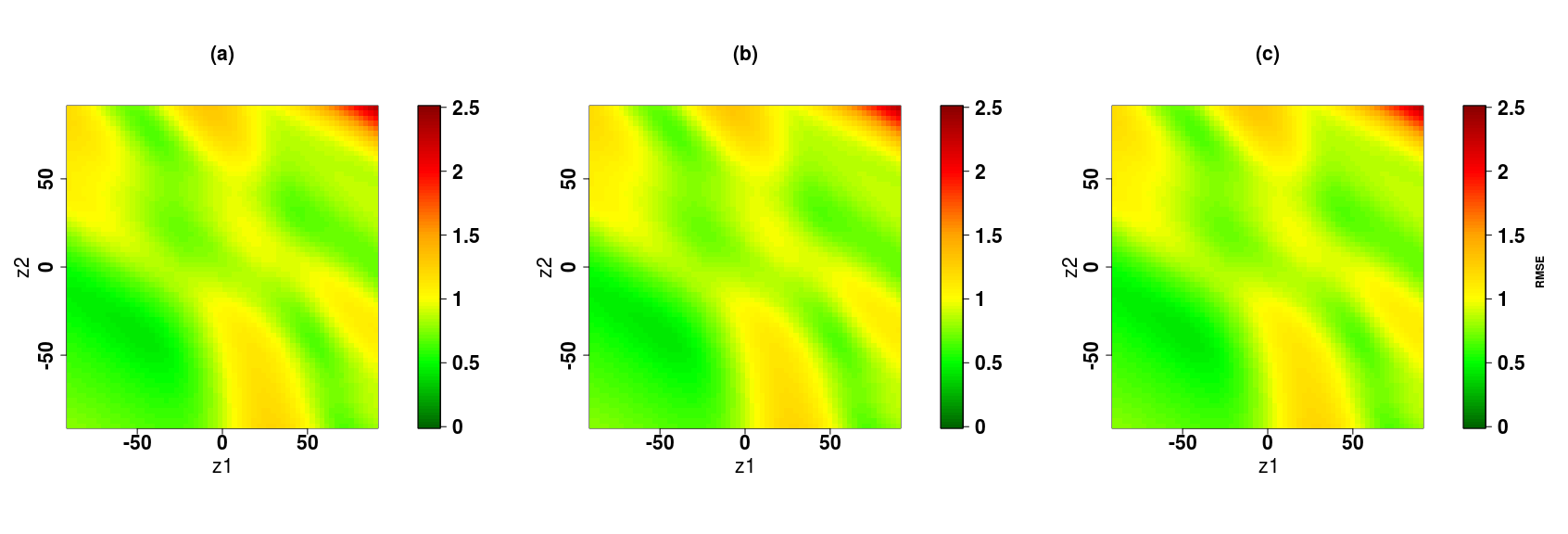}
    \caption{The RMSE maps obtained by $\textnormal{GP}^{\textnormal{FPCA}}_{\textnormal{wavelet}}$, $\textnormal{GP}^{\textnormal{FPCA}}_{\textnormal{B-splines}}$, and  $\textnormal{GP}^{\textnormal{PCA}}$ which are respectively named (a), (b), and (c).}
    \label{fig:RmseCampbell2D}
\end{figure}

In this section, we build a test sample with $n_{test} = 1000$ simulations of $f$. The inputs $\mathbf{x}$ are drawn at random independently from the uniform distribution on $[-1, 5]^8$.
The output maps are assumed to be unknown. They are estimated by $\textnormal{GP}^{\textnormal{PCA}}$, $\textnormal{GP}^{\textnormal{FPCA}}_{\textnormal{wavelet}}$, or $\textnormal{GP}^{\textnormal{FPCA}}_{\textnormal{B-splines}}$, using the parameters chosen in section \ref{sec:TestCase:subsec:calibration}, and based on the $n=200$ learning samples. The root mean square error (Eq. \ref{eq:RMSEtest}) of each method is compared in Figure \ref{fig:RmseCampbell2D}.

\begin{equation}
\textnormal{RMSE}(\mathbf{z}) = \sqrt{\frac{1}{n_{test}}\sum_{i'=1}^{n_{test}} \left[y_{\mathbf{x}^{(i')}}(\mathbf{z})-\hat{y}_{\mathbf{x}^{(i')}}(\mathbf{z})\right]^2},\ \ \mathbf{z}\in [-90,90]^2
\label{eq:RMSEtest}
\end{equation}where $y_{\mathbf{x}^{(i')}}(.)$ and $\hat{y}_{\mathbf{x}^{(i')}}(.)$  are respectively the true and predicted output map for the input $\mathbf{x}^{(i')}$, with $i'=1,\dots,n_{test}$. 

We can see that the three methods have the same prediction accuracy.

\review{The prediction accuracy can be also quantified by another criterion, called the $Q^2$ criterion. A generalized version (from \cite{marrel2010global}) has been used:}
\begin{equation}
\label{eq::Q2}
    Q^2 = 1-\frac{\mathbb{E}_\mathbf{z}\left\{\mathbb{E}_\mathbf{X}\left[\left(Y_\mathbf{X}(\mathbf{z})-\hat{Y}_\mathbf{X}(\mathbf{z})\right)^2\right]\right\}}{\mathbb{E}_\mathbf{z}\left\{\mathbb{V}\textnormal{ar}_\mathbf{X}\left[Y_\mathbf{X}(\mathbf{z})\right]\right\}} =
    1 - \frac{\mathbb{E}_\mathbf{z}\left\{\textnormal{MSE}(\mathbf{z})\right\}}{\mathbb{E}_\mathbf{z}\left\{\mathbb{V}\textnormal{ar}_\mathbf{X}\left[Y_\mathbf{X}(\mathbf{z})\right]\right\}}.
\end{equation}
In practice, expectations are replaced by empirical means. The $Q^2$ criterion compares the MSE errors of the model relatively to the variance of observations, averaged spatially. When $Q^2$ is greater than 0, the model performs better than predicting by the mean of observations, and the closest to 1 the better is prediction accuracy.
Here, $\textnormal{GP}^{\textnormal{FPCA}}_{\textnormal{wavelet}}$, $\textnormal{GP}^{\textnormal{FPCA}}_{\textnormal{B-splines}}$, and $\textnormal{GP}^{\textnormal{PCA}}$, all have a predictability coefficient $Q^2 \approx 96.6\%$, which is very satisfactory.

It can be concluded that the three metamodels are equally efficient. 
This is encouraging for the FPCA-based techniques, which seem to be a good competitor to PCA on this difficult analytical function, while reducing significantly the problem dimension. Indeed,
$\textnormal{GP}^{\textnormal{FPCA}}_{\textnormal{wavelet}}$ uses only $29,3\%$ of wavelet coefficients. $\textnormal{GP}^{\textnormal{FPCA}}_{\textnormal{B-splines}}$ reduces first the dimension to $1225$ instead of $4096$. The interest in terms of computational time is not representative for this analytical case. It will be visible on the real case application where the dimension of the model output is larger (Section \ref{sec:StudyCase}).



 \subsection{\review{Variant without orthonormalization}}
 
 \review{The version of FPCA without orthonormalization of the B-splines basis has also been performed. Thus, the selection step of the procedure has been modified by using a Lasso regression model as described in  
 Section \ref{sec:metamodelsFPCA}. We have used a common penalty parameter $\lambda$ for all $\mathbf{x}$, 
 in order to save time in computations.}\\
 
\begin{figure}[H]
    \centering
    \includegraphics[width=0.9\textwidth]{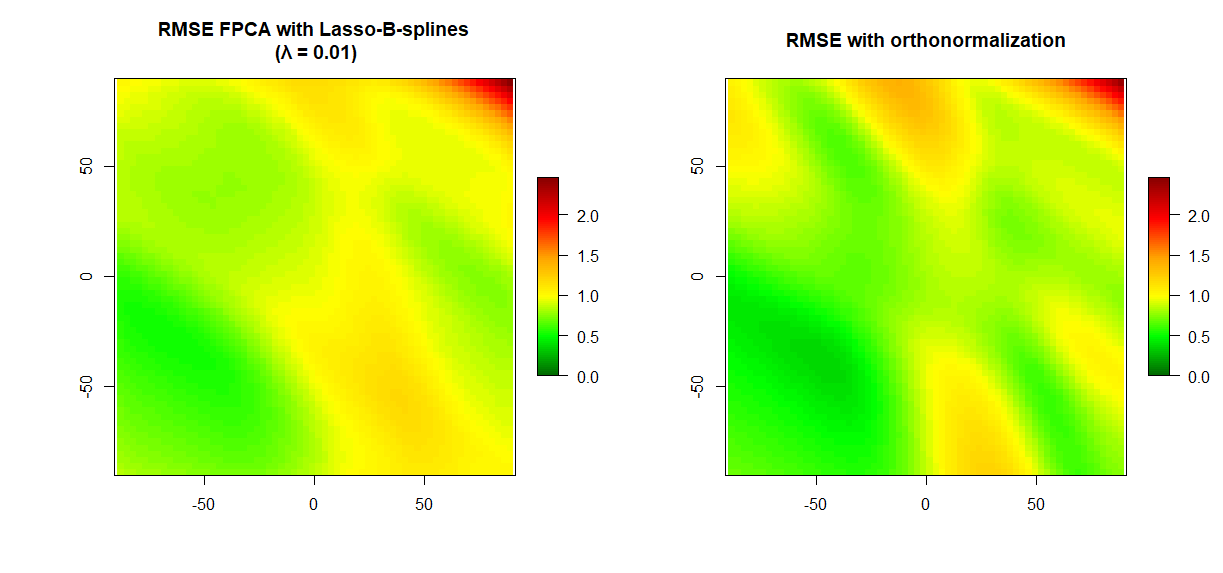}
    \caption{\review{RMSE maps : on the left, without orthonormalization by using Lasso regression, on the right, by using $\textnormal{GP}^{FPCA}_{B-splines}$.}}
    \label{fig:LassoComparison}
\end{figure}

 \review{FPCA parameters have been chosen by applying the same procedure as in Section \ref{sec:TestCase:subsec:calibration}. In addition to $\tilde{K}$ and $n_{PC}$, the penalty parameter $\lambda$  (Eq. \ref{eq::Lasso}) has been tuned. 
 The final selected value
 is $\lambda=0.01$, resulting in the
 lowest RMSE values for $\tilde{K}=1225$ and $n_{PC}=5$. 
 We notice that $\tilde{K}$ and $n_{PC}$ have the same values as $\textnormal{GP}^{FPCA}_{B-splines}$.}\\
\review{
The RMSE map has then been computed for the test set, 
by applying the same procedure as in Section \ref{sec:TestCase:subsec:prediction}. It is compared to the one obtained with $\textnormal{GP}^{FPCA}_{B-splines}$ on Figure \ref{fig:LassoComparison}.
We can see that $\textnormal{GP}^{FPCA}_{B-splines}$, constructed from the mainline version of Algorithm \ref{algo:metamodelsFPCA}, remains slightly better in terms of prediction accuracy.}

\review{The codes for both FPCA were executed on a single core of an AMD $\textrm{Ryzen}^\textrm{TM}$ 7 4700U CPU. The computational time for the variant of our FPCA algorithm based on Lasso is greater than one minute, compared to less than one second for FPCA with B-splines orthonormalization.
Due to the dimensionality of the maps in the next case study (flooding maps), the variant of FPCA with Lasso is intractable. 
Therefore, that variant will not be used in the rest of the paper.}

\subsection{Global sensitivity analysis}
\label{sec:TestCase:subsec:GSI}

We now perform a global sensitivity analysis of Campbell 2D function, based on metamodelling. 
Following the results of the previous section, all three metamodels are very accurate, and we will use the $\textnormal{GP}^{\textnormal{FPCA}}_{\textnormal{B-spline}}$ metamodel. 
In section \ref{sec:background:subsec:GSI}, a generalized sensitivity index has been defined. 
Property \ref{prop:GSI_PCA} indicates that GSI are equal to the average of Sobol indices of principal components, weighted by eigenvalues. Therefore, GSI estimation directly relies on the estimation of Sobol indices. Here, we have used the estimator defined in \cite{saltelli2002making}, depending on two samples.
Hence, two input sample sets of size $n_0=10^4$ have been randomly generated, which imposes a total of $n_0(d+2)=10^5$ model runs. The initial sample sets are Latin Hypercube Samples (LHS), drawn at random from the uniform distribution. 

\begin{figure}[H]
    \centering
    \includegraphics[width=0.8\textwidth]{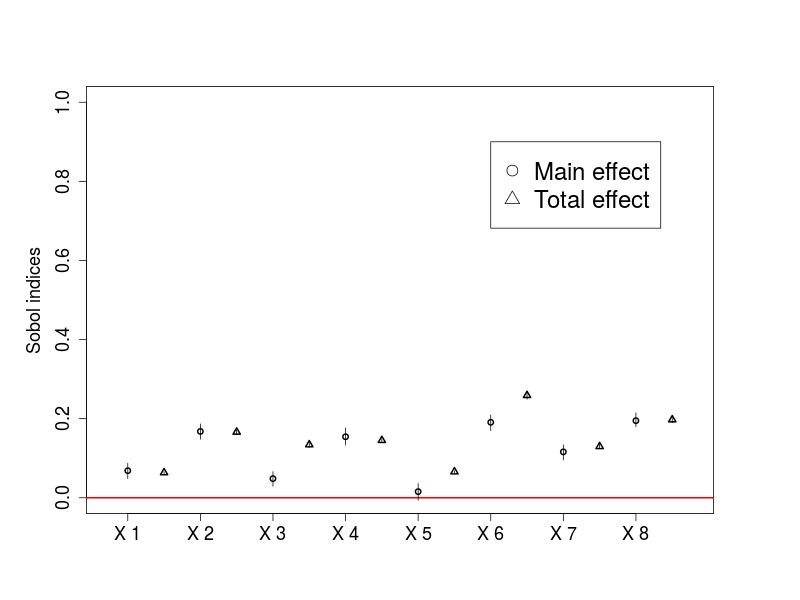}
    \caption{Generalized sensitivity indices (GSI) estimations of the $8$ input variables (first order with circle points, and total order with triangle points).}
    \label{fig:GsiCampbell}
\end{figure}

Figure \ref{fig:GsiCampbell} shows the estimations of the generalized sensitivity indices (GSI). $X_6$ is the most influential with the highest total order index. $X_8$ is the second most influential input with a main effect equal to the one of $X_6$, and a lower total effect. We notice that its influence is entirely defined by its main effect (total and first indices are equal). $X_2$, $X_4$ and $X_7$ are also entirely defined by their main effect. They corresponds to the third, fourth and fifth influential input variables. $X_1$, $X_3$ and $X_5$ are the three lowest influential variable (with the respective order). $X_1$ is entirely defined by its main effect. Finally, $X_3$ and $X_5$ are mainly influential in interaction with other variables (small total indices and negligible values of first order indices).

\section{Application on coastal flooding model}
\label{sec:StudyCase}


\subsection{Description of the case study}
\label{sec:TestCaseFlooding}
\begin{figure}[H]
\begin{center}
\includegraphics[width=0.7\textwidth]{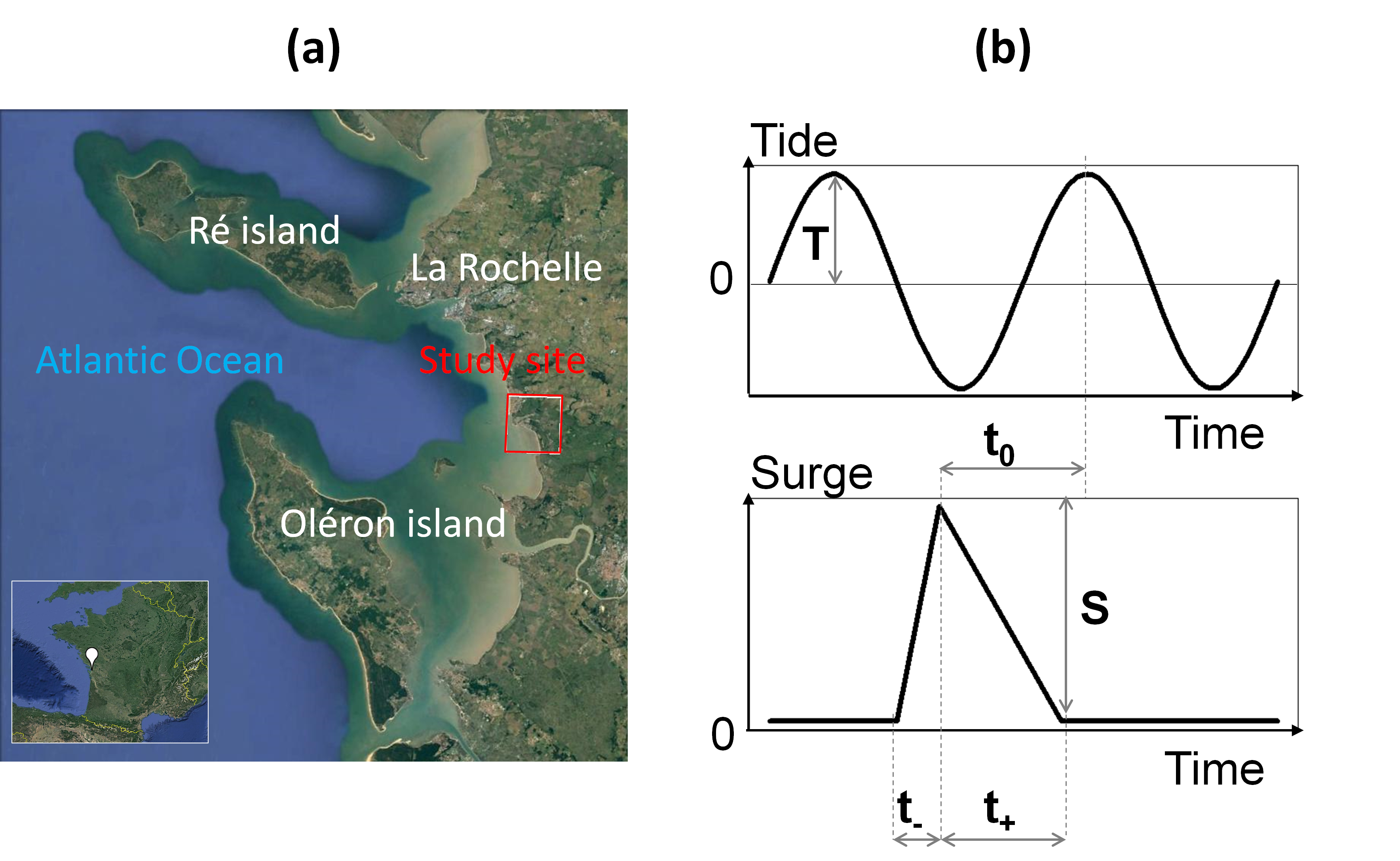}
\end{center}
\caption{a) Site location, b) Parameterization of time evolution of tide and surge}
\label{fig:CaseStudy}
\end{figure}

The methodology in section \ref{sec:TestCase} is also applied on a case of coastal flooding. The study site is ``Les Bouchôleurs'' (french Atlantic coast, near ``La Rochelle'' city), which was hit during the Xynthia storm in 2010 (see Figure \ref{fig:CaseStudy}). The main flooding processes correspond to overflow.\\

Coastal flooding processes are simulated with the numerical code MARS \cite{lazure2008external}, where adaptations were made by the BRGM to take into account specificities of local flooding processes (hydraulic processes around connections like nozzles, spillways, etc. and breaching phenomena) \cite{rohmer2018casting}.\\

\begin{figure}[H]
\begin{center}
\includegraphics[width=\textwidth]{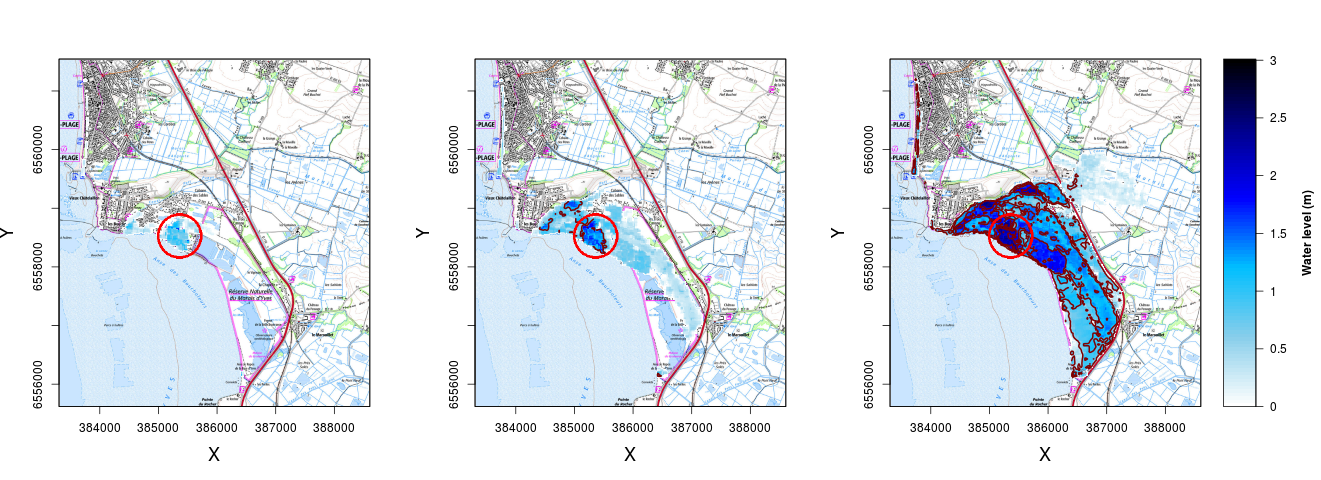}

\caption{From left to right, spatial outputs of the coastal flooding numerical model for three input configurations, namely $\mathbf{x_1}=$(3.61 m, 1.75 m, 5.72 hours, -3.10 hours, 2.11 hours), $\mathbf{x_2}=$(3.51 m, 1.68 m, 3.93 hours, -5.82 hours, 5.85 hours),  and $\mathbf{x_3}=$(3.23 m, 1.55 m, 0.19 hours, -3.66 hours, 3.06 hours). The red circle corresponds to the location of the main urban area. Brown lines of the middle and right maps are contour lines of water levels. The background layer (SCAN 25® from the National Institute of Geographic and Forest Information IGN) indicates the locations of the urban areas and key topographic elements (roads, railways, marshlands, etc.).}
\label{fig:CoastalModelOutput}
\end{center}
\end{figure}

We focus on the interplay between tide and storm surge on the spatial distribution of the maximum water depth after flooding. Here, the time evolution of both signals is simplified: tide is assimilated to a sinusoidal curve with $T$ the high tide level (between $0.95$m and $3.70$m); the storm surge is assumed to be triangular (see figure \ref{fig:CaseStudy}) using four parameters, namely $S$ the peak amplitude (ranging between $0.65$m and $2$m), $t_0$ the phase difference between surge peak and high tide (between $-6$ an $6$ hours), $t_+$ and $t_{-}$ the time duration of the increase and the decrease of the storm surge signal (between $0.5$ and $12$ hours). We are interested in the sensitivity to the $5$ input parameters $\mathbf{x}=(T,S,t_0,t_+,t_{-})$.
The output of the simulator corresponds to a map with \review{regular discretizations of} $256\times 256$ pixels (each pixel being of $25\times25m^{2}$). For example, Fig. \ref{fig:CoastalModelOutput} shows three output maps (the darker the blue is, the higher the water level is) considering three input configurations.\\

Depending on inputs' values, the flooding spatial extent is more or less important (Figure \ref{fig:CoastalModelOutput}). First, we notice structural infrastructures constraining the flood: the main local road (in black) and the national road (in red). Both roads (more ever the national road), being built slightly higher than the surrounding (on embankments) to avoid road flooding, limits the penetration of water inland. However, it does not act completely as a dike  as it is not impermeable (existence of hydraulic  connections between the east and west areas of the road). Second, we notice sharp irregularities of the water level in the red circle area, especially in the middle map of Figure \ref{fig:CoastalModelOutput}. This area corresponds to the location of the main urban area on the study site. Furthermore, dark blue pixels are located in the vicinity of light blue pixels colors (borders are delimited with brown line in the middle and right maps of Figure \ref{fig:CoastalModelOutput}). This means that the water level can strongly vary from one pixel to another. We can see on Fig.\ref{fig:CoastalModelOutput}-right, a border delimited by dark blue area and a lighter blue one: these abrupt changes can be explained by the transition between different types of land cover (as shown on the background layer of Figure \ref{fig:CoastalModelOutput}), i.e. different Manning coefficients, which influence water propagation; for instance, from urban to rural zone. In cities, it can come from different types of structural components like road layout, bridges, succession of buildings, their heights etc. These examples illustrate the complexity and heterogeneity of flooding maps.\\ 


Because of the computation time cost of the simulator ($\approx$ $0.5$ to $1$ hour for one simulation), only a limited number of simulations ($n=500$) were performed by randomly choosing configurations of $\mathbf{x}$ using a Sobol random sequence (see e.g., \cite{bratley1988algorithm}).

\subsection{Prediction accuracy}
\label{sec:StudyCase:subsec:prediction}

The three metamodelling methods presented in section \ref{sec:TestCase} are also compared in the coastal flooding case. The dataset of simulation results  contains $253$ flooded maps, and $247$ maps without any flooding (i.e. all water depths are at zeros). Metamodels have been trained using a learning dataset of $n_{learning} = 400$ maps. Half of them have been randomly chosen among runs for which flooding has occurred (the other half thus corresponding to maps without any flooding). In order to test the metamodels' prediction accuracy, we use the remaining $n_{test}=100$ ones as test samples.\\ 
The settings of FPCA-based methods have been done as in Section 5.
We have chosen D4 Daubechies wavelets with one level of resolution. We have used B-splines of degree 1, 
with $100$ knots equally spaced on each dimension of the spatial domain.
Based on the $10$-fold cross-validation results, we have chosen $n_{PC} = 2$ for all three methods. \review{The two first principal components correspond to  $96\%$ of the explained inertia for all three.} We have chosen a total of $\review{\tilde{K}}=4\,000$ and $\review{\tilde{K}}=1\,700$ coefficients respectively for the PCA step in $\textnormal{GP}^{FPCA}_{wavelet}$ and $\textnormal{GP}^{FPCA}_{B-spline}$ methods. This corresponds to a reduction of respectively $94\%$ and $97\%$ in terms of number of variables, compared to standard PCA which works on the whole vector of $256^2=65 \,536$ pixels. 

\begin{figure}[H]
    \centering
    \includegraphics[width=0.49\textwidth]{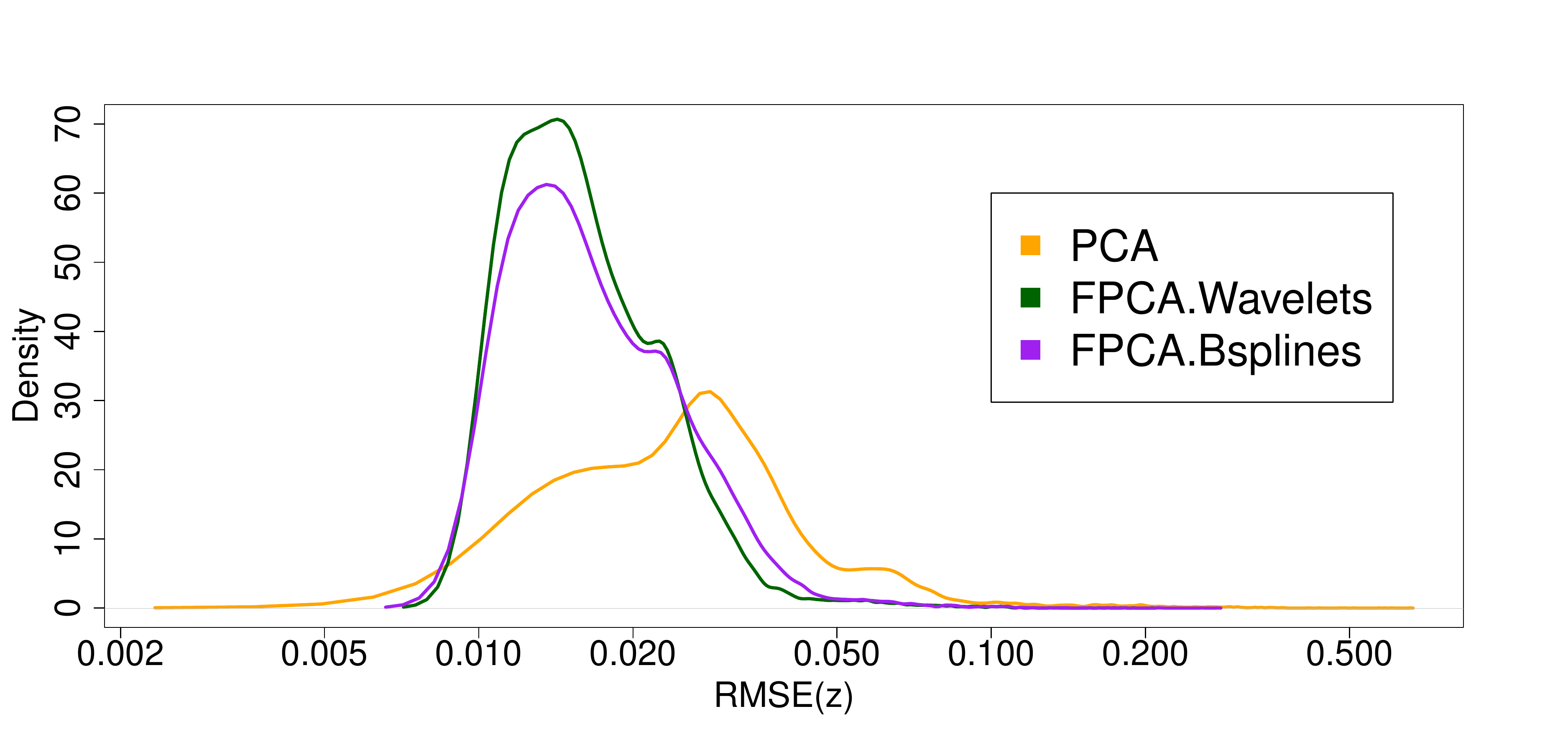}
    \includegraphics[width=0.49\textwidth]{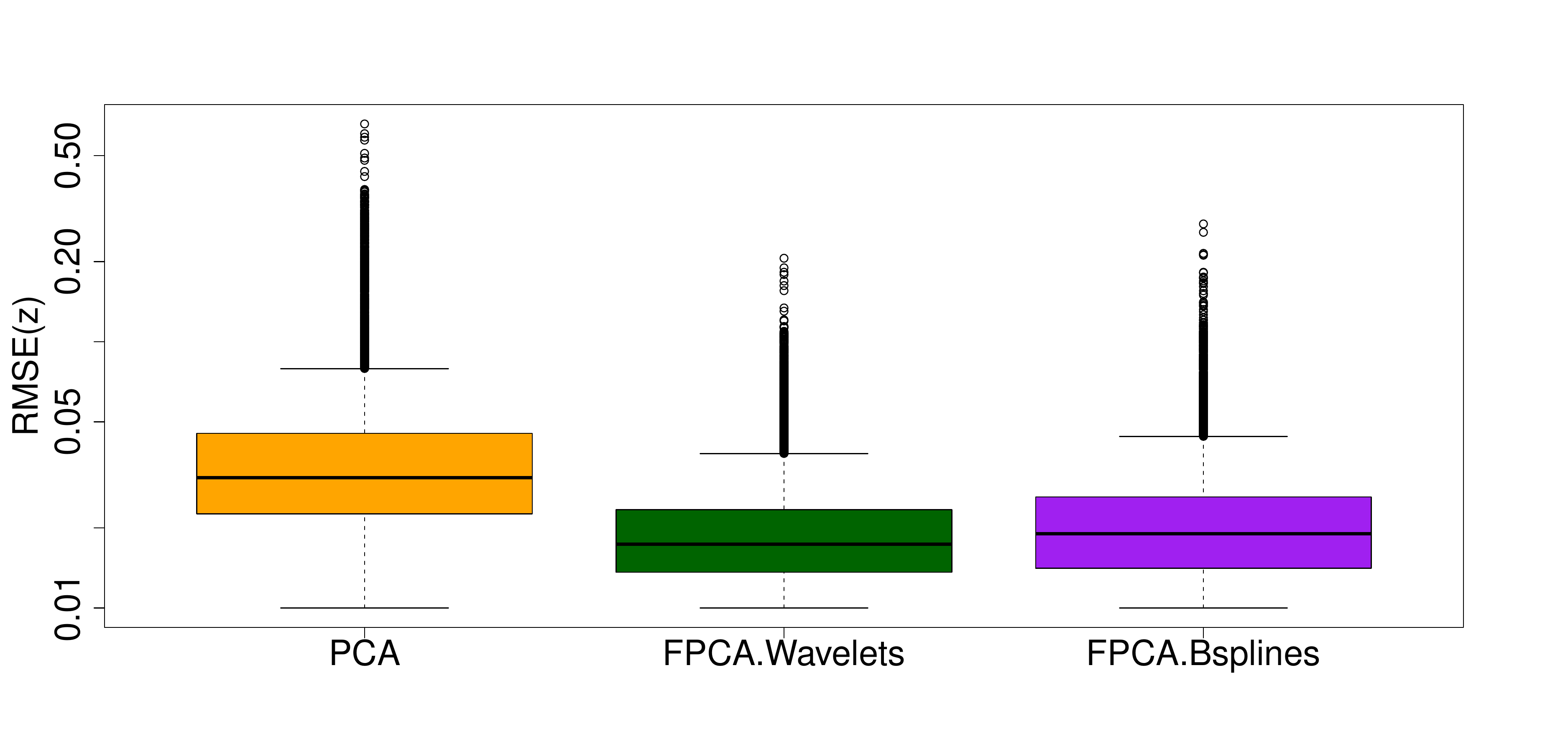}
    \caption{Density (Left) and Boxplots (Right) of spatial RMSE (expressed in meters). RMSE are plotted in log base 10 scale. The points of the right figure represent RMSE values which are outside the whiskers defined as 1.5 times the interquartile range from the box.}
    \label{fig:DensityBoxplotBrgm}
\end{figure}

We first compare the performance of the three methods globally by analyzing the distribution of spatial errors, measured by RMSE$(\mathbf{z})$, for the whole map. \review{Notice that RMSE is preferred to $Q^2$ here. Indeed, it is expressed in meters and thus easily interpretable in a risk assessment perspective. Furthermore, the $Q^2$ criterion cannot be computed for the pixels corresponding to unflooded area as the denominator is equal to zero}. Boxplots and estimated probability density functions are shown in Figure \ref{fig:DensityBoxplotBrgm}, in log scale. 
Looking at these errors, we can see that the two FPCA-based methods 
$(\textnormal{GP}^{FPCA}_{wavelet}, 
\textnormal{GP}^{FPCA}_{B-spline})$
outperform the PCA-based one ($\textnormal{GP}^{PCA}$), both on average and for extreme values. Thus, the mode, the median and the third quartile are clearly smaller for FPCA methods. Furthermore, extreme values (visible on boxplots) are limited to 0.2 m for FPCA methods, contrarily to PCA for which they can reach they can reach 0.5 m. Finally, the FPCA method based on wavelets 
is slightly more accurate here.

\begin{figure}[H]
    \centering
    \includegraphics[width=\textwidth]{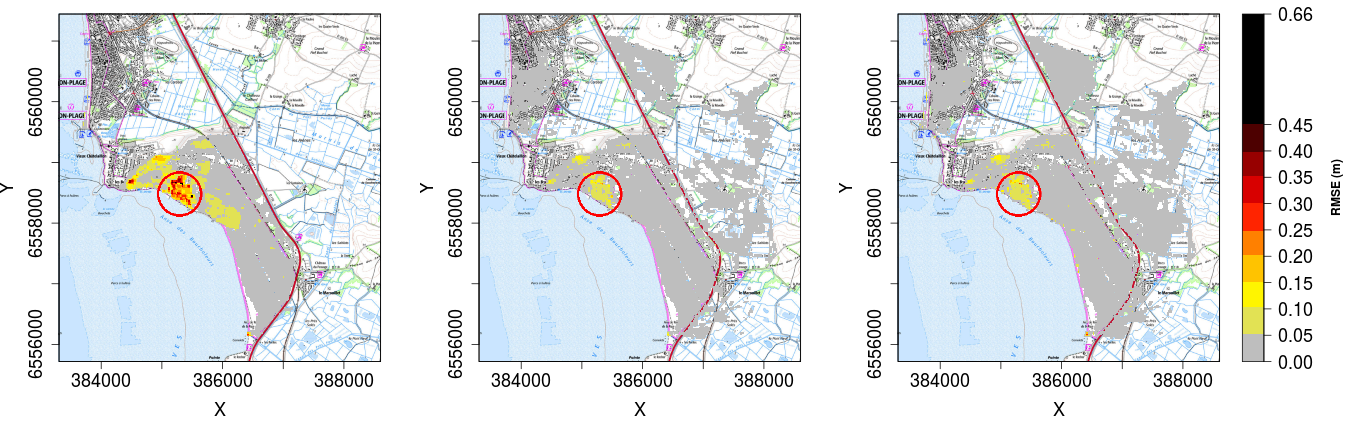}
    \caption{The RMSE maps (using 100 test samples) obtained with $\textnormal{GP}^{PCA}$, $\textnormal{GP}^{FPCA}_{wavelet}$, and $\textnormal{GP}^{FPCA}_{B-spline}$ for the coastal flooding case. In the maps, there are locations without any given values. They correspond to locations where RMSE is strictly less than $1$cm, which is negligible. We use the same  background layer as for Figure \ref{fig:CoastalModelOutput}}
    \label{fig:RmseBrgm}
\end{figure}



Densities and boxplots in Figure \ref{fig:DensityBoxplotBrgm} give a global spatial information about prediction accuracy of the three methods. The advantages of FPCA approach are also analyzed locally in Figure \ref{fig:RmseBrgm}, which compares spatial RMSE obtained with the 3 methods. For $\textnormal{GP}^{PCA}$, highest errors can be noticed where irregularities are observed in Figure \ref{fig:CoastalModelOutput}, i.e. in the urban area (outlined by a red circle), where there is a  spatial heterogeneity. In these areas, $\textnormal{GP}^{FPCA}_{wavelet}$ and $\textnormal{GP}^{FPCA}_{B-spline}$ RMSE are lower than $\textnormal{GP}^{PCA}$ RMSE by $0.10$ m to $0.20$ m. 
However, outside this central zone, RMSE values of both FPCA methods appear to be slightly higher than using PCA by no more than $0.05$ m, which is a reasonable order of magnitude.\\

\review{In the next section,} we will use  $\textnormal{GP}^{FPCA}_{wavelet}$ to perform a sensitivity analysis of the coastal flooding model\review{, because of its better accuracy}. \\

\subsection{Sensitivity analysis}

As in section \ref{sec:TestCase:subsec:GSI}, a sensitivity analysis has been performed for the coastal flooding model, by replacing the simulator by the metamodel (combined with $\textnormal{GP}^{FPCA}_{wavelet}$) trained with $n=500$ simulations. The method of section \ref{sec:TestCase:subsec:GSI} has been used with 
$n_0=10^4$ Monte-Carlo random samples and assuming uniform law for each input (over their respective range of variation). 

\begin{figure}[H]
    \centering
    \includegraphics[width=0.8\textwidth]{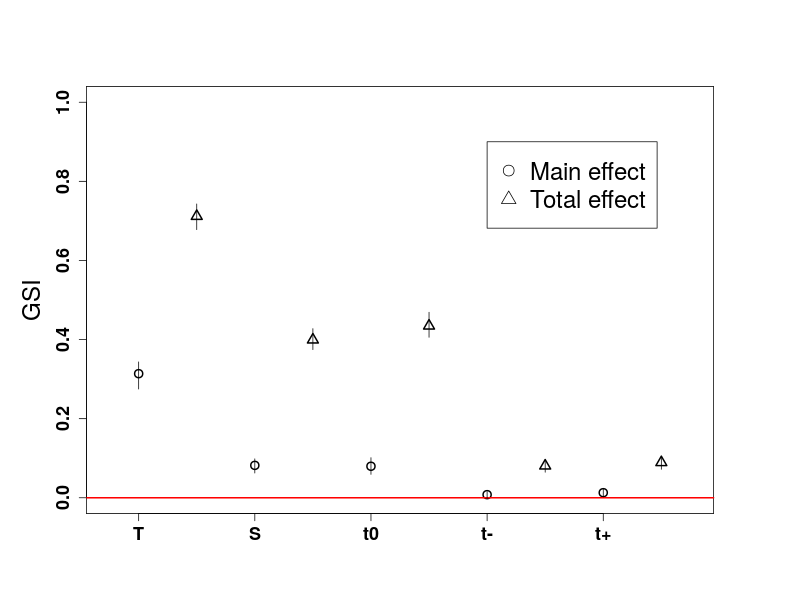}
    \caption{Generalized sensitivity indices of coastal flooding model, which measure influence of sea forcing parameters. The main effects of the input variables are illustrated by circles points. The total effects are illustrated by triangular points.}
    \label{fig:GsiBrgm}
\end{figure}

The estimated generalized sensitivity indices are shown in Figure \ref{fig:GsiBrgm}. The tide level $T$ appears to have the highest influence, as indicated by the large first-order Sobol index. The difference between the main and total effects shows that $T$ has strong interaction with the other input variables. The other two most influential variables (of same importance) are the surge $S$ and the phase difference $t_0$. They are mainly influential in interaction with other variables (the first order indices are approximately $0.1$, instead of $0.4$ for the total indices). The two remaining variables, $t_{-}$ and $t_{+}$, have negligible effect, with a total effect of around $0.1$ for both. This result appears to be physically consistent with the overflowing processes in this zone, which are mainly caused by the maximum water level (i.e. related to the $T$, $S$ and $t_0$) reached offshore.  
Finally, these results validate the relevance of this metamodelling approach for sensitivity analysis. 


%
%
 
\section{Conclusion and future works}
\label{sec:conclusion}
In this paper, we introduce a methodology combining meta-modelling and sensitivity analysis for models with high-dimensional spatial output including strong discontinuities. This work was motivated by the sensitivity analysis of a coastal flooding model. 

To this aim, we propose to combine metamodels with functional principal component analysis (FPCA) to reduce the dimension of the spatial output, i.e. to combine the advantages of functional basis approximation and of PCA dimension reduction. 
\review{To further reduce dimension, we have added a preliminary selection step.
That selection can be done either directly on the basis with a penalized regression approach, or with an energy criterion after orthonormalization. That second approach was the most successful in our experiments, both in terms of accuracy and computational cost, and presents the advantage of providing a physical interpretation as well.}

Two types of basis have been compared: wavelets and B-splines. First, the methodology has been tested on an analytical test case where FPCA gives the same results as PCA approach. 
This shows that there is no loss of accuracy when performing two nested decomposition for FPCA. The interest of the methodology has then been analyzed on a real case of coastal flooding. Our experimental results show that FPCA meta-modelling approach is more accurate than PCA for the estimation of water levels in areas where sharp irregularities are present. 
Coastal flooding maps used for this paper, are matrices of dimension $256\times 256$: this enabled us to conduct PCA and to compare the results with FPCA. In practice, higher dimensions (for which PCA is hardly feasible) can be considered with our approach, even if B-splines basis are used. In addition, sensitivity analysis is performed  using an extended formulation of generalized sensitivity indices that are valid to any basis functions avoiding the assumption of orthonormality. The application on the real case of these indices allows identifying inputs in agreement with the overflowing processes in this zone.\\ 

Several lines of improvement have been identified. Firstly, predicting whether or not flooding occurs is still challenging,
although the predicted water depth is small in absence of flooding. 
This may be related to some threshold effects that control coastal processes. If the water level at the coast (which results from storm surge and tide characteristics) is lower than a specified threshold, flooding cannot occur: the water height at any given location inland remains zero. Otherwise, provided that the water level slightly increases and exceeds a specified threshold, overflow-induced inundation can occur and
inland locations may be flooded. To tackle this effect, the following 
potential solutions should be explored: classification method in order to learn inputs where there is any inundation (see an example by \cite{rohmer2018casting}), or by adding constraints on the GP metamodels \cite{AndresPhD2019}.\\         
Secondly, although the usage of a functional basis aims at preserving spatial regularity, 
some flooded areas, in grey, are not always enough connected together in the predicted maps and consequently, connected to the sea. 
However, in the physical model, flow propagation comes from the sea and flooded areas are always continuous, unless the model represents hydraulic connections, such as nozzle.
The problem may be addressed by adding a global regularity criterion to the energy criterion used to select basis coefficients. \review{In this vein, the alternative Lasso criterion used in the selection step could be investigated further.}\\
Thirdly, sensitivity indices have been estimated using the variance as a measure of uncertainty. This might not be adapted to represent physic phenomenon with threshold effect (which may induce some multi-modality in the output probability distribution), as it is the case for coastal flooding. Future work should then consider alternative uncertainty measures (like dependence measure \cite{da2015global},\cite{de2017sensitivity}).\\

\section{Acknowledgement}

\review{We are grateful to an associate editor and two anonymous reviewers for their relevant comments, 
that led to substantial improvements.}
The project is funded by the French geological survey BRGM and the French state reinsurance company CCR. The authors thank for their feedbacks the participants of the Chair in Applied Mathematics OQUAIDO, gathering partners in technological research (BRGM, CEA, IFPEN, IRSN, Safran, Storengy) and academia (CNRS, Ecole Centrale de Lyon, Mines Saint-Etienne, University of Grenoble, University of Nice, University of Toulouse) around advanced methods for Computer Experiments.

\bibliography{mybibfile}

\begin{thebibliography}{10}
\expandafter\ifx\csname url\endcsname\relax
  \def\url#1{\texttt{#1}}\fi
\expandafter\ifx\csname urlprefix\endcsname\relax\def\urlprefix{URL }\fi
\expandafter\ifx\csname href\endcsname\relax
  \def\href#1#2{#2} \def\path#1{#1}\fi

\bibitem{chaumillon2017storm}
E.~Chaumillon, X.~Bertin, A.~B. Fortunato, M.~Bajo, J.-L. Schneider,
  L.~Dezileau, J.~P. Walsh, A.~Michelot, E.~Chauveau, A.~Cr{\'e}ach, et~al.,
  Storm-induced marine flooding: Lessons from a multidisciplinary approach,
  Earth-science reviews 165 (2017) 151--184.

\bibitem{naulin2016estimation}
J.~Naulin, D.~Moncoulon, S.~Le~Roy, R.~Pedreros, D.~Idier, C.~Oliveros,
  Estimation of insurance related losses resulting from coastal flooding in
  {F}rance, Natural Hazards and Earth System Sciences Discussions 3 (2015)
  2811--2846.

\bibitem{Xynthia2011}
FFSA, Gema, La tempête {X}ynthia du 28 février 2010. {B}ilan chiffré au 31
  décembre 2010, Tech. rep., Association française de l'assurance (02 2011).

\bibitem{iooss2015review}
B.~Iooss, P.~Lema{\^\i}tre, A review on global sensitivity analysis methods,
  in: Uncertainty management in simulation-optimization of complex systems,
  Springer, 2015, pp. 101--122.

\bibitem{saltelli2008global}
A.~Saltelli, M.~Ratto, T.~Andres, F.~Campolongo, J.~Cariboni, D.~Gatelli,
  M.~Saisana, S.~Tarantola, Global sensitivity analysis: the primer, John Wiley
  \& Sons, 2008.

\bibitem{chen2011efficient}
T.~Chen, K.~Hadinoto, W.~Yan, Y.~Ma, Efficient meta-modelling of complex
  process simulations with time--space-dependent outputs, Computers \& chemical
  engineering 35~(3) (2011) 502--509.

\bibitem{marrel2010global}
A.~Marrel, B.~Iooss, M.~Jullien, B.~Laurent, E.~Volkova, Global sensitivity
  analysis for models with spatialy dependent outputs, Environmetrics 22~(3)
  (2010) 383--397.

\bibitem{jia2013}
G.~Jia, A.~A. Taflanidis, Kriging metamodeling for approximation of
  high-dimensional wave and surge responses in real-time storm/hurricane risk
  assessment, Computer Methods in Applied Mechanics and Engineering 261 (2013)
  24--38.

\bibitem{marrel2015}
A.~Marrel, N.~Perot, C.~Mottet, Development of a surrogate model and
  sensitivity analysis for spatio-temporal numerical simulators, Stochastic
  environmental research and risk assessment 29~(3) (2015) 959--974.

\bibitem{li2020}
M.~Li, R.-Q. Wang, G.~Jia, Efficient dimension reduction and surrogate-based
  sensitivity analysis for expensive models with high-dimensional outputs,
  Reliability Engineering \& System Safety 195 (2020) 106725.

\bibitem{Ma2019}
P.~Ma, A.~Mondal, B.~Konomi, J.~Hobbs, J.~Song, E.~Kang, Computer model
  emulation with high-dimensional functional output in large-scale observing
  system uncertainty experiments, arXiv preprint arXiv:1911.09274.

\bibitem{williams2006gaussian}
C.~K.~I. Williams, C.~E. Rasmussen, Gaussian processes for machine learning,
  MIT Press Cambridge, MA, 2006.

\bibitem{ramsay2006functional}
J.~O. Ramsay, Functional data analysis, Wiley Online Library, 2006.

\bibitem{lamboni2011multivariate}
M.~Lamboni, H.~Monod, D.~Makowski, Multivariate sensitivity analysis to measure
  global contribution of input factors in dynamic models, Reliability
  Engineering \& System Safety 96~(4) (2011) 450--459.

\bibitem{Mar2015}
A.~Marrel, N.~Perot, C.~Mottet, Development of a surrogate model and
  sensitivity analysis for spatio-temporal numerical simulators, Stochastic
  environmental research and risk assessment 29~(3) (2015) 959--974.

\bibitem{redd2012comment}
A.~Redd, A comment on the orthogonalization of b-spline basis functions and
  their derivatives, Statistics and Computing 22~(1) (2012) 251--257.

\bibitem{mallat1999wavelet}
S.~Mallat, A wavelet tour of signal processing, Elsevier, 1999.

\bibitem{daubechies1988orthonormal}
I.~Daubechies, Orthonormal bases of compactly supported wavelets,
  Communications on pure and applied mathematics 41~(7) (1988) 909--996.

\bibitem{sobol1993sensitivity}
I.~M. Sobol, Sensitivity estimates for nonlinear mathematical models,
  Mathematical modelling and computational experiments 1~(4) (1993) 407--414.

\bibitem{homma1996importance}
T.~Homma, A.~Saltelli, Importance measures in global sensitivity analysis of
  nonlinear models, Reliability Engineering \& System Safety 52~(1) (1996)
  1--17.

\bibitem{gamboa2014sensitivity}
F.~Gamboa, A.~Janon, T.~Klein, A.~Lagnoux, et~al., Sensitivity analysis for
  multidimensional and functional outputs, Electronic Journal of Statistics
  8~(1) (2014) 575--603.

\bibitem{bjorck1994numerics}
{\AA}.~Bj{\"o}rck, Numerics of gram-schmidt orthogonalization, Linear Algebra
  and Its Applications 197 (1994) 297--316.

\bibitem{qin2000general}
K.~Qin, General matrix representations for b-splines, The Visual Computer
  16~(3-4) (2000) 177--186.

\bibitem{liuob}
X.~Liu, H.~Nassar, K.~Podg{\'O}rski, Splinets--efficient orthonormalization of
  the b-splines, arXiv preprint arXiv:1910.07341.

\bibitem{tibshirani1996regression}
R.~Tibshirani, Regression shrinkage and selection via the lasso, Journal of the
  Royal Statistical Society: Series B (Methodological) 58~(1) (1996) 267--288.

\bibitem{ManuelR}
{R Core Team}, \href{https://www.R-project.org/}{R: A Language and Environment
  for Statistical Computing}, R Foundation for Statistical Computing, Vienna,
  Austria (2018).
\newline\urlprefix\url{https://www.R-project.org/}

\bibitem{package}
The \textnormal{GpOutput2D} package,
  \url{https://github.com/tranvivielodie/GpOutput2D}.

\bibitem{dupuy2015dicedesign}
D.~Dupuy, C.~Helbert, J.~Franco, et~al., Dicedesign and diceeval: Two r
  packages for design and analysis of computer experiments, Journal of
  Statistical Software 65~(11) (2015) 1--38.

\bibitem{hastie2009cross}
T.~Hastie, R.~Tibshirani, J.~Friedman, Cross-validation, 241--249). New York,
  NY, USA: Springer New York Inc.

\bibitem{saltelli2002making}
A.~Saltelli, Making best use of model evaluations to compute sensitivity
  indices, Computer physics communications 145~(2) (2002) 280--297.

\bibitem{lazure2008external}
P.~Lazure, F.~Dumas, An external--internal mode coupling for a 3d
  hydrodynamical model for applications at regional scale (mars), Advances in
  water resources 31~(2) (2008) 233--250.

\bibitem{rohmer2018casting}
J.~Rohmer, D.~Idier, F.~Paris, R.~Pedreros, J.~Louisor, Casting light on
  forcing and breaching scenarios that lead to marine inundation: Combining
  numerical simulations with a random-forest classification approach,
  Environmental Modelling \& Software 104 (2018) 64--80.

\bibitem{bratley1988algorithm}
P.~Bratley, B.~L. Fox, Algorithm 659: Implementing {S}obol's quasirandom
  sequence generator, ACM Transactions on Mathematical Software (TOMS) 14~(1)
  (1988) 88--100.

\bibitem{AndresPhD2019}
A.~F. López-Lopera, Gaussian process modelling under inequality constraints,
  Ph.D. thesis, Mines Saint-Etienne (2019).

\bibitem{da2015global}
S.~Da~Veiga, Global sensitivity analysis with dependence measures, Journal of
  Statistical Computation and Simulation 85~(7) (2015) 1283--1305.

\bibitem{de2017sensitivity}
M.~De~Lozzo, A.~Marrel, Sensitivity analysis with dependence and variance-based
  measures for spatio-temporal numerical simulators, Stochastic environmental
  research and risk assessment 31~(6) (2017) 1437--1453.

\end{thebibliography}

\end{document}